\documentclass[fleqn,usenatbib]{mnras}

\usepackage{newtxtext,newtxmath}

\usepackage[T1]{fontenc}

\DeclareRobustCommand{\VAN}[3]{#2}
\let\VANthebibliography\thebibliography
\def\thebibliography{\DeclareRobustCommand{\VAN}[3]{##3}\VANthebibliography}

\usepackage{graphicx}	
\usepackage{amsmath}	

\newcommand{\lya}{\rm\,Ly$\alpha$}
\newcommand{\ha}{\rm\,H$\alpha$}
\newcommand{\hb}{\rm\,H$\beta$}
\newcommand{\oiii}{\rm\,[O{\sc iii}]}

\newcommand{\nii}{\rm\,[N{\sc ii}]}

\def\hi{{\rm\,H{\sc i}}}

\title[Filamentary protocluster traced by triple NB]
{Association of cold gas, massive galaxies, and AGNs in a filamentary protocluster traced by triple narrow-band imaging}

\author[K. Daikuhara et al.]{Kazuki Daikuhara,$^{1,2}$\thanks{E-mail: daikuhara@ir.isas.jaxa.jp}
Tadayuki Kodama,$^{2}$
Haruka Kusakabe,$^{3,4}$
Charles C. Steidel,$^{5}$
Ichi Tanaka,$^{6}$
\newauthor
Satoshi Kikuta,$^{7}$
Hideki Umehata,$^{8,9}$
Rhythm Shimakawa,$^{10}$
Yusei Koyama,$^{3}$
Kentaro Motohara,$^{11,12}$
\newauthor
Masahiro Konishi,$^{11}$
Jose Manuel Perez Martinez,$^{13,14}$
Mariko Kubo,$^{2,15}$
Dawn Erb,$^{16}$
Kosuke Takahashi,$^{2}$
\newauthor
and Keita Fukushima.$^{17}$
\\
\\
$^{1}$Institute of Space and Astronautical Science, Japan Aerospace Exploration Agency, 3-1-1, Yoshinodai, Chuou-ku, Sagamihara, Kanagawa 252-5210, Japan\\
$^{2}$Astronomical Institute, Tohoku University, 6-3, Aramaki, Aoba, Sendai, Miyagi, 980-8578, Japan\\
$^{3}$National Astronomical Observatory of Japan (NAOJ), National Institutes of Natural Sciences (NINS), 2-21-1 Osawa, Mitaka, Tokyo 181-8588, Japan\\
$^{4}$Department of General Systems Studies, Graduate School of Arts and Sciences, The University of Tokyo, 3-8-1 Komaba, Meguro-ku, Tokyo, 153-8902, Japan\\
$^{5}$Cahill Center for Astronomy and Astrophysics, California Institute of Technology, MS 249-17, Pasadena, CA 91125.\\
$^{6}$Subaru Telescope, National Astronomical Observatory of Japan, 650 North A’ohoku Place, Hilo, HI 96720, USA\\
$^{7}$Department of Astronomy, School of Science, The University of Tokyo, 7-3-1 Hongo, Bunkyo, Tokyo 113-0033, Japan\\
$^{8}$Institute for Advanced Research, Nagoya University, Furocho, Chikusa, Nagoya 464-8602, Japan\\
$^{9}$Department of Physics, Graduate School of Science, Nagoya University, Furocho, Chikusa, Nagoya 464-8602, Japan\\
$^{10}$Waseda Institute for Advanced Study (WIAS), Waseda University, 1-21-1, Nishi-Waseda, Shinjuku, Tokyo 169-0051, Japan\\
$^{11}$Advanced Technology Center, National Astronomical Observatory of Japan, 2-21-1 Osawa, Mitaka, Tokyo 181-8588, Japan\\
$^{12}$Institute of Astronomy, Graduate School of Science, The University of Tokyo, 2-21-1 Osawa, Mitaka, Tokyo 181-0015, Japan\\
$^{13}$Instituto de Astrofísica de Canarias (IAC), E-38205, La Laguna, Tenerife, Spain.\\
$^{14}$Universidad de La Laguna, Departamento Astrofísica, E-38206 La Laguna, Tenerife, Spain\\
$^{15}$School of Science, Kwansei Gakuin University, Sanda, Hyogo 669-1337, Japan\\
$^{16}$The Leonard E. Parker Center for Gravitation, Cosmology and Astrophysics, Department of Physics, University of Wisconsin-Milwaukee,\\ 3135 N. Maryland Avenue, Milwaukee, WI 53211, USA\\
$^{17}$Institute for Data Innovation in Science, Seoul National University, Seoul 08826, Republic of Korea\\
}

\date{Accepted 2025 October 13. Received 2025 October 04; in original form 2025 May 08}

\pubyear{2025}

\begin{document}
\label{firstpage}
\pagerange{\pageref{firstpage}--\pageref{lastpage}}
\maketitle

\begin{abstract}
We investigate galaxy populations in the HS1700+64 protocluster at $z=2.30$, characterized by two prominent linear filaments traced by spatially extended \lya\ blobs. We conducted a wide area mapping of emission line galaxies across the protocluster using the unique combination of three matched narrow-band filters, corresponding to \lya, \ha, and \oiii\ emission lines at $z=2.30$. We find that H$\alpha$ emitters are strongly clustered at the intersection of the filaments, suggesting a protocluster core. In contrast,  \lya\ emitters tend to avoid the dense region and the filaments, likely due to the resonant scattering of \lya\ photons by \hi\ gas and/or enhanced dust attenuation in galaxies associated with these structures. These findings support a scenario in which cold gas flows via filaments and to the core, fed by the cold-stream mode accretion in the early phase of protocluster assembly, and promoting active star formation there. Further evidence of the scenario comes from the alignment of massive, evolved galaxies in those filaments traced by distant red galaxies, suggesting accelerated galaxy growth in the filaments in the early Universe. This study clearly shows observationally that accelerated galaxy formation takes place not only in the protocluster core but also in the associated surrounding filamentary structure. This underscores the critical role of large-scale filaments in efficiently accumulating the cold gas and channeling it to galaxies therein and to the protocluster core. Such vigorous gas assembly facilitates star formation activity and drives galaxy growth in the early stage of cluster formation.
\end{abstract}




\section{Introduction}\label{sec:Introduction}

Galaxy clusters are the most massive gravitationally bound, kinematically relaxed systems in the present-day Universe. Their progenitors, protoclusters, are expected to develop at the intersections of large-scale filamentary structures at high redshifts, where cold gas is efficiently accreted via cold streams without being heated up to the virial temperature of the dark matter halo (cold stream mode). This process enables efficient star formation. 
As the halo mass grows, the intracluster gas is eventually shock heated to high temperature and fully ionized, thereby suppressing the supply of cold gas to the galaxies within the halo \citep[hot mode;][]{Dekel2009}.
The transition phase from the cold stream mode to the hot mode is predicted to occur around $z \sim 2$, when the physical conditions of the protoclusters significantly change \citep{Dekel2009,Daddi2022}.
This epoch, known as “cosmic noon,” represents the peak period of the cosmic star formation rate (SFR) density \citep[e.g.,][]{Madau2014}.
Semi-analytic model or cosmological simulation indicate that high-redshift protoclusters contribute to cosmic SFR density but galaxy clusters at $z=0$ contribute negligibly to it \citep[e.g.,][]{Chiang2017,Fukushima2023}.
These results underscore the importance of studying the protoclusters in the early Universe.

Moreover, cosmological simulations indicate that protocluster outskirts also contribute significantly to the total cosmic SFR density \citep{Fukushima2023}.  
Galaxies in the present-day clusters are assembled from many extended regions along the surrounding large-scale structures \citep{Keres2005, Dekel2009, Cautun2014}.
Within the filamentary structures on the outskirts, there are localized moderate overdense regions such as groups.
Galaxies in the surrounding filaments or groups may experience some early environmental effects before they are assembled to cluster cores ("pre-processing") \citep[e.g.,][]{Zabludoff1998,Fujita2004,Chiang2017}.  
Observational evidence for pre-processing in infalling groups and filaments has been reported in nearby cluster samples \citep{Haines2015, Sarron2019}.

Pre-processing can alter the star formation activities, gas content, and chemical properties of galaxies.
Large surveys further show that, in the nearby universe ($z \lesssim 0.1$--0.5), proximity to filaments correlates with star formation \citep{Chen2017,Kuutma2017,Malavasi2022}.
This phase encompasses a variety of physical processes -- galaxy-galaxy interactions/mergers, efficient cold gas accretion, strangulation (gas reservoir stripping), and so on. -- which can vary significantly depending on the structures and local environmental conditions of galaxies in the outskirts of protoclusters. 

In some cases, galaxies near filaments show evidence for \hi\ gas supply, consistent with continued accretion from the intrafilament medium \citep{Kleiner2017}, whereas other studies find \hi\ deficiency near filaments at fixed local density and stellar mass \citep{CroneOdekon2018}. 
The latter suggests that cold gas accretion has been suppressed, and these contrasting results may reflect differences in gas supply mechanisms. 
Large-scale structures are therefore not merely regions of enhanced density, but can directly regulate the \hi\ content of galaxies, playing a critical role in their evolutionary pathways.

In order to understand the early environmental effects, it is thus important to fully cover all various environments along the large-scale structures, and investigate how galaxy properties are dependent on different environments.
Therefore, it is crucial, first of all, to clearly distinguish among different environments around protoclusters, such as cores, outskirts, groups, and filaments.

Filamentary structures connecting surrounding groups and cluster cores through which galaxies are assembling are of particular interest \citep[e.g.,][]{Cautun2014}.
We do not know how those relatively primitive environment could affect galaxy properties already there in the early phase of pre-processing.
However, identifying such relatively underdense, intermediate-density regions is difficult because of weaker contrast against projections and contaminations of galaxies.
Moreover, tracing cold \hi\ gas associated with the filaments is even more challenging, especially at high redshifts.
The gas is traced only through the indirect method, such as \lya\ absorption lines embedded in the spectra of background sources.
Recent Ly$\alpha$-forest tomography with dense background sightlines demonstrates feasibility of mapping the IGM at $z\sim2$ -- 3 \citep{Lee2014a, Lee2014b,Newman2022,Newman2025}.
However, the limited sight lines to bright background objects result in poor spatial resolution and thus hamper us from identifying detailed environments such as filaments.

One of the key questions is whether the observed filamentaly structure seen as the sequence of galaxies also hosts abundant cold gas, as the efficient gas provider to galaxies in the filaments and the protoclusters cores at their intersections.
Such gas association along the structures would be strongly connected to the early environmental dependence of galaxy properties.

\lya\ blobs (LABs) may serve as valuable tracers of associated cold gas trapped in potential wells of massive galaxies or group-scale haloes \citep{Hayes2011,Dijkstra2009,Rosdahl2012,Daddi2022}.
LABs are among the brightest LAEs with spatially extended diffuse \lya\ emission, often spanning tens of kpc to 100 kpc \citep[e.g.,][]{Steidel2000}.
LABs appear in relatively overdense regions, suggesting their possible physical link to protoclusters and large-scale structures \citep{Steidel2000,Matsuda2004,Matsuda2011,Erb2011,Mawatari2012,Prescott2008,Yang2010,Badescu2017,Kikuta2019,Huang2022,Zhang2023,Ramakrishnan2023}. 
According to \cite{Yuan2014}, a 1 $h^{-1}$ Gpc cosmological N-body simulation suggests that large, bright LABs may be associated with massive halos, placing them at the higher end of the halo mass distribution, and therefore they are likely to assemble and end up in the cores of present-day galaxy clusters. 
Although the origin of LABs remains uncertain, possible origins related to \hi\ gas include resonant scattering \citep{Hayes2011} and cold gas accretion \citep{Dijkstra2009,Rosdahl2012,Daddi2022}.
Since gas accretion is crucial in regulating star formation both in entire protoclusters and within individual member galaxies, understanding the properties and distribution of cold gas is essential for unveiling the environmental dependence of star formation and chemical evolution. In this context, LABs may serve as valuable probes of these cold gas–related processes in the early Universe.

Although searching for spatially extended diffuse \lya\ emission (i.e., \lya\ halos/blobs) around galaxies is an effective technique to trace associated cold gas, its application is limited by severe cosmological surface brightness dimming at high redshifts. 
\cite{Shimakawa2017b} propose a novel approach, namely, \lya–\ha\ dual narrow-band (NB) imaging. 
Since \lya\ line is a resonant line that is readily scattered by the surrounding \hi\ gas, while \ha\ is not.
Therefore, the \lya/\ha\ line ratio measured within a given aperture -- or the relative number density of \lya\ emitters (LAEs) versus \ha\ emitters (HAEs) -- can serve as an indicator of the escape fraction of \lya\ photons and thus the presence of \hi\ gas. 
We note that dust attenuation is more severe for \lya\ (rest-frame 1216 \AA) than for \ha\ (6563 \AA), so the \lya/\ha\ line ratio is also affected by dust attenuation.
To identify the presence of \hi\ gas from \lya/\ha\ line ratio, we need to properly correct for dust attenuation.

This study first aims to validate this technique and investigate how \hi\ gas is distributed across different environments (cores and filaments).
Next, we aim to elucidate the interrelationships among gas inflows/outflows, star formation, and the surrounding environment (filaments versus cores). In particular, establishing the relationship between the distribution of \hi\ gas and SF activities within protoclusters is crucial for identifying early environmental effects. For example, efficient gas supply along surrounding filaments may be linked to accelerated galaxy growth within protoclusters, which is manifested by enhanced SFRs. Narrow-band imaging has been widely employed to efficiently study the environmental dependence of star formation compared with spectroscopy \citep[e.g., MAHALO-Subaru survey:][]{koyama2010,koyama2011,koyama2013,koyama2014,Tanaka2011,tadaki2012,Hayashi2016,Shimakawa2018uss,Shimakawa2018pks,Daikuhara2024}. The young, clumpy protocluster USS1558–003 (hereafter USS1558) at $z=2.53$ exhibits both enhanced star formation and a metallicity deficit \citep{Daikuhara2024,Jose2024}. These phenomena are likely driven by cold gas accretion as well as galaxy mergers or interactions. 
By combining various NB imaging analyses, we aim to shed light on how cold gas fuels galaxy evolution in the early Universe.

This paper is organized as follows: Section~\ref{sec:target} explains our target, Section~\ref{sec:Data} describes data reduction and source detection, and Section~\ref{sec:Selection} details the sample selection. Section~\ref{sec:resutls} presents results from NB analysis, spatial distribution, and SF activity. We propose a dual NB technique to photometrically select active galactic nuclei (AGNs) in Section~\ref{subsec:AGN}. A discussion follows in Section~\ref{sec:discussion}, with conclusions summarized in Section~\ref{sec:summary}.

In this paper, we assume a cosmology with $H_0 = 70$ km s$^{-1}$ Mpc$^{-1}$, $\Omega_M = 0.3$, and $\Omega_\Lambda = 0.7$. At the protocluster redshift of $z=2.3$, 1 arcmin corresponds to a physical scale of 462 kpc or a comoving scale of 1.52 Mpc. All magnitudes are presented in the AB system \citep{Oke1983}. Stellar masses and SFRs are estimated assuming a Chabrier initial mass function (IMF) \citep{Chabrier2003}.

\section{Target protocluster}\label{sec:target}

Our target is the HS1700+64 protocluster (hereafter HS1700) located at $z=2.30$. 
This structure was originally identified as an overdensity of UV/optically-selected SF galaxies at $z = 2.30$ in the sight line towards a quasar HS~1700+6416 at $z=2.74$ across a sky area of $15.3\times15.3$ arcmin \citep{Steidel2004,Steidel2005}.
It is one of the best-studied protoclusters, exhibiting a redshift-space overdensity of $\delta \sim 7$, suggesting that it will evolve into a rich cluster with a halo mass of $M_h \sim 10^{15}\, \mathrm{M_{\odot}}$ by the present-day \citep{Steidel2005}. 
A \lya\ NB imaging survey has revealed six large \lya\ blobs with sizes exceeding 100 kpc and \lya\ luminosities of $\sim10^{43}$ erg s$^{-1}$ \citep{Bogosavljevic2010,Steidel2011,Erb2011}. 
Intriguingly, these seven LABs are nearly perfectly aligned along two linear filaments spanning at least 12 co-moving Mpc, suggesting a link between the physical origin of LABs and that of linear filamentary structures \citep[Umehata et al. in prep;][]{Bogosavljevic2010,Erb2011}.
In this protocluster, \cite{Digby-North2010} identified X-ray AGN fraction for BX/MD galaxies is $6.9^{+9.2}_{-4.4}$\% using a $\sim$200 ks Chandra/ACIS-I observation, suggesting a preferential AGN activity in dense environments at $z>2$.
Herschel/SPIRE \citep{Kato2016} and SCUBA-2 850 $\mu$m \citep{Lacaille2019} observations show the active the dusty star-formation activity.
\citet{Kato2016} reported the inferred SFR densities are significantly higher ($\times\sim10^{4}$) compared to the star-formation activity in the field.
They selected dusty star-forming galaxies based on the colours (S350$_{\mathrm{\mu m}}$/S250$_{\mathrm{\mu m}}$ and S500$_{\mathrm{\mu m}}$/S350$_{\mathrm{\mu m}}$) using Herschel/Spectral and Photometric Imaging Receiver (SPIRE) survey.
These studies suggest that HS1700 is exhibiting active star formation and AGN activity, and is likely in an active phase of mass accumulation.

\section{Data acquisition and reduction}\label{sec:Data}


\begin{table*}
    \centering
    \caption{Specifications of the narrow-band filters. From left to right, filter name, instrument, target line, central redshift, central wavelength and width of each filter, redshift range, velocity range, and line of sight co-moving distance that each filter corresponds to, and reference, are shown in the table.}
    \label{tab:filter}
    \begin{tabular}{ccccccccc}\hline
    Filter & Instrument & Target line & Redshift & $\lambda_{\mathrm{centre}}$ & $\Delta \lambda_{\mathrm{filter}}$ & $\Delta z$ & $\Delta v$  & Publication\\
    &&&&[$\mu$m]&[$\mu$m] && [km]  \\
    \hline
    Br$\gamma$ & MOIRCS & \ha\   & 2.30 & 2.1650 & 0.025 & 0.04 & 3364 &  This work \\
    NB2167     & SWIMS  & \ha\   & 2.30 & 2.1640 & 0.022 & 0.03 & 3462 &  This work \\
    NB1653     & SWIMS  & \oiii\ & 2.30 & 1.6520 & 0.016 & 0.03 & 3048 &  This work\\
    NB4010     & LFC Wide-Field Imager & \lya\  & 2.30 & 0.4010  & 0.009 & 0.07 & 2904 &  \citet{Bogosavljevic2010,Erb2011}\\
    &&&&&&&&\citet{Steidel2011}\\
    \hline
    \end{tabular}
\end{table*}

\subsection{Photometric dataset}

\begin{figure}
\begin{center} 
\includegraphics[width=\columnwidth]{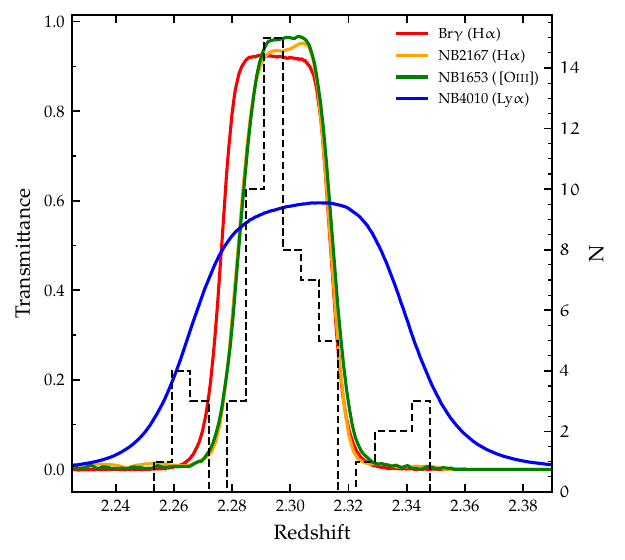}
\end{center} 
\caption{Filter response functions in redshift space of each NB filter; MOIRCS/Br$\gamma$ (red), SWIMS/NB2167 (orange), SWIMS/NB1653 (green), and Palomar/NB4010 (blue). The dashed histogram represents the redshift distribution of the spectroscopically confirmed cluster member galaxies identified by Keck Baryonic Structure Survey (KBSS) \citep{Rudie2012,Steidel2014,Strom2017}.} 
\label{fig:filter}
\end{figure}

We utilize the photometric dataset of HS1700 protocluster that consists of Un, G, and Rs (William Hershel Telescope) \citep{Steidel2004}, 24$\mu$m (MIPS/Spitzer) \citep{Reddy2010}, NB4010 (LFC Wide-Field Imager/Hale 200-inch telescope at Palomar Observatory) \citep{Steidel2011,Erb2011}, J, J1, J2, H, K, NB1653, NB2167 (SWIMS/Subaru; this work), J, H, Ks, and Br$\gamma$ (MOIRCS/Subaru; this work).
MOIRCS and SWIMS instruments stand for Multi-Object Infra-Red Camera and Spectrograph \citep[MOIRCS;][]{Ichikawa2006, suzuki2008} and Simultaneous-color Wide-field Infrared Multi-object Spectrograph \citep[SWIMS;][]{Konishi2012,Konishi2018,Konishi2020,Motohara2014,Motohara2016}, respectively.
All four NB filters that we used in the analyses (Table~\ref{tab:filter}) are properly matched in redshift space to the corresponding line for the HS1700 protocluster members (Figure~\ref{fig:filter}).

\subsection{MOIRCS}\label{subsec:MOIRCS}
We conducted Br$\gamma$, J, H, Ks imaging on 2020 July 2 (Observatory Staff Time for Ichi Tanaka), May 4, 5, 6, 7 (S23A-017; PI Tadayuki Kodama), and May 27 (S23A-101S  H. Kusakabe, S23A-011; PI Zhaoran Liu) on Subaru Telescope.
We observed with three pointings, where the seeing ranged from 0.55 to 0.62 arcsec.
We convolved these images using a Gaussian kernel to match the
optical images, resulting in the final seeing size of 1.0 arcsec.
The MOIRCS Br$\gamma$ filter ($\lambda=2.165$ $\mu$m, FWHM = 0.025 $\mu$m) can capture \ha\ emission line from galaxies in the redshift slice of $z=2.30\pm0.02$.
Data reduction was performed with the imaging data pipeline MCSRED\footnote{https://www.naoj.org/staff/ichi/MCSRED/mcsred.html} \citep{Tanaka2011}.
Reduction processes include bias subtraction, flat fielding, sky subtraction, mask creation, coadding, distortion correction, and mosaicing.
Astrometry is performed using SCAMP \citep{Bertin2006} based on Gaia-EDR3 \citep{Gaia2016,Gaia2021}.
MOIRCS has two HAWAII-2 arrays (2028 $\times$ 2028 pixels), which have a slight difference in sensitivity.
We correct for this by adopting different zero-points from the standard star observations.
Finally, The pixel scales of the images are matched to the existing optical images (0.236 arcsec per pixel).

\subsection{SWIMS}\label{subsec:SWIMS}
SWIMS is capable of simultaneous two-color imaging ("blue" at 0.9 $\mu$m -- 1.45 $\mu$m and "red" at 1.45 $\mu$m -- 2.5 $\mu$m) with a field of view of 6.6$^{\prime}$ $\times$ 3.3$^{\prime}$ (in the case of mounted on Subaru Telescope). 
We observed with NB1653, NB2167, J, J1, J2, H, and Ks imaging on 2022 April 12, 13, 14, 15, 16 (S22A-031; PI H. Kusakabe) on Subaru Telescope.
We conducted two pointing observations, where the seeing ranged from 0.65 to 0.71 arcsec.
We convolved these images using a Gaussian kernel to match the optical images, resulting in the final seeing size of 1.0 arcsec.
The pair NB filters on SWIMS, namely, NB2167 ($\lambda=2.164$ $\mu$m, FWHM = 0.022 $\mu$m) and NB1653 ($\lambda=1.652$ $\mu$m, FWHM = 0.016 $\mu$m) correspond to \ha\ and \oiii\ emission lines, respectively, from galaxies exactly in the same redshift slice of $z=2.30\pm0.02$ by design.

Data reduction was performed with the imaging data pipeline SWSRED\footnote{https://www.ioa.s.u-tokyo.ac.jp/TAO/swims/?Data\_Reduction} \citep{Konishi2020}.
Reduction processes include bias subtraction, flat fielding, sky subtraction, mask creation, coadding, distortion correction, mosaicing, and astrometry.

In addition to the normal reduction processes implemented in the pipeline, we conduct two more reduction processes on the SWIMS data. 
We independently subtract the global background, which the pipeline cannot handle, and the striped background derived from the 64-channel wide electrical signal.
First, we use the object mask created by the pipeline to estimate the global background with {\it photutils.background.Background2D} \citep{Photutils}. 
A kernel is created with a radial basis function kernel overlaid with white noise in 64 $\times$ 64-pixel boxes. 
We then use {\it scikit-learn} \citep[ver. 1.6.1;][]{scikitlearn} to estimate and subtract the background for the images cut out for each 64-channel section.

Since the persistence appears in all bands (from Y to K-band), we identified bright stars in each frame and masked them with a square mask for ten consecutive frames taken before and after. Subsequently, we apply sigma clipping to the sigma map, and if any persistence signal ($>$ 5 sigma) remains after masking, additional masking is performed. This method allows us to remove the persistence signal from all frames.
It is particularly noticeable in the blue band. In some cases, note that it is detected as false objects.

Finally, the images are matched in pixel scale to the existing optical images (0.236 arcsec per pixel).

\subsection{Triple narrow-band imaging}
The most unique feature of this study is the coordinated emission line galaxies survey across the protocluster with triple narrow-band (NB) imaging. In this work, we conduct dual NB imaging with MOIRCS and SWIMS on Subaru for HAEs and \oiii\ emitters (O3Es), respectively, and combine them with the existing \lya\ NB imaging on LFC on Palomar \citep{Erb2011}.
The combined NB imaging technique is very effective in tracing real structures of physically associated member galaxies because the multiple detections of emission lines with multiple NB filters ensure their redshifts (Figure~\ref{fig:filter}).
Moreover, this method can provide us with emission line ratios using only imaging observations, which can then provide us with various information about galaxy properties such as the association of \hi\ gas (Section~\ref{subsec:filament}) and the presence of AGNs (Section~\ref{subsec:AGN}).

\begin{table}
    \centering
    \caption{Average 5$\sigma$ limiting magnitude of our data set. These values are measured with a random 1.2 arcsec diameter aperture photometry in blank regions. Note that since the mosaic image is composed of our data with varying depths, the values should be regarded as just average value.}
    \label{tab:filter2}
    \begin{tabular}{cccc}\hline
    \ \ \ \ &\ \ \ \ Filter\ \ \ \  &\ \  Instrument\ \  & Ave. 5$\sigma$ limiting mag.\\
    \hline
    NB & NB1653 (\oiii)    & SWIMS  & 23.24\\
    NB & Br$\gamma$ (\ha) & MOIRCS & 23.22\\
    NB & NB2167 (\ha) & SWIMS & 22.95\\
    MB & J1 & SWIMS & 24.38\\
    MB & J2 & SWIMS & 23.92\\
    BB & Un & WHT & 26.86\\
    BB &G & WHT & 27.22\\
    BB &Rs & WHT & 26.44\\
    BB &J & MOIRCS & 24.10\\
    BB &J & SWIMS & 23.92\\
    BB &H & MOIRCS & 23.62\\
    BB &H & SWIMS & 24.00\\
    BB &Ks & MOIRCS & 23.71\\
    BB &Ks & SWIMS & 23.79\\
    \hline
    \end{tabular}
\end{table}

\subsection{Source detection}\label{subsec:Detection}

We perform source extractions from the reduced co-added NB images with SExtractor \citep[version 2.25.2,][]{Bertin1996}.
We also carry out photometries with a 1.5 arcsec diameter aperture and the Kron radius using the double-imaging mode of the SExtractor \citep{Bertin1996}.
We use the NB image as the detection image.
The SExtractor parameters for detections/photometries were set to \textit{DETECT\_MINAREA = 9, DETECT\_THRESH = 1.2, ANALYSIS\_THRESH = 1.2, DEBLEN\_MINCONT = 0.001, BACK\_SIZE= 64, BACK\_FILTERSIZE = 5, BACKPHOTO\_TYPE = LOCAL, BACKPHOTO\_THICK = 32}, and $k=2.5$, following \cite{Daikuhara2024}.
We conducted source detections in the mosaiced image with varying depths.
To account for spatial variations in background noise, we evaluate the noise level based on the weight image following the method of \cite{Shimakawa2024a}.
Background noise ($\sigma$) is estimated by random photometry with Photoutils \citep{Photutils} following \cite{Bunker1995,Shimakawa2018pks,Shimakawa2024a,Daikuhara2024}.
Initially, we conduct random aperture photometry using circular apertures with diameters ranging from 0.5 to 2.0 arcseconds, incrementing by 0.1 arcsec steps.
The aperture photometry and the average weight within each aperture are determined using the {\it Photoutils} package \citep{Photutils}.
The photometry results are then binned into five groups based on the weight values ($W$), and the parameter $C_1$ in the equation $\sigma(W)= C_1\, W^{-0.5}$ is determined by fitting.
Subsequently, the average weight of each target aperture is substituted into the fitted formula, allowing us to derive the parameters $C_2$ and $N$ in the equation $\sigma(S)_{W=w}= C_2\, S^N$.
These formula enables us to compute photometric errors for any aperture size $S$ given a specific weight $w$. 
Finally, using the actual aperture areas of the targets, we estimate their respective photometric errors.
The results for all detected objects are presented in Figures~\ref{fig:err_1} --  Figure~\ref{fig:err_6}.

\section{Sample selection}\label{sec:Selection}

\subsection{Colour -- magnitude diagram}\label{sec:Colourmag}
\begin{figure*}
\begin{minipage}{5.6cm} 
\begin{center} 
\includegraphics[width=5.6cm]{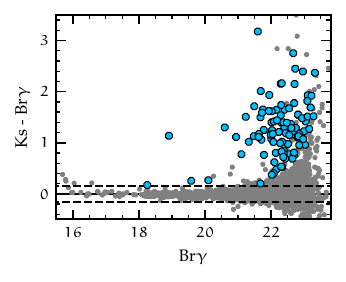}
\end{center} 
\end{minipage} 
\begin{minipage}{5.6cm} 
\begin{center} 
\includegraphics[width=5.6cm]{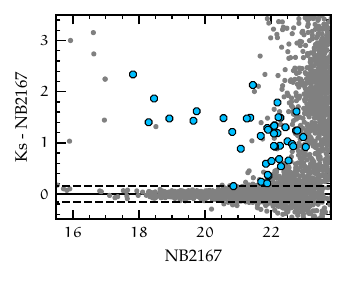}
\end{center} 
\end{minipage}
\begin{minipage}{5.6cm} 
\begin{center} 
\includegraphics[width=5.6cm]{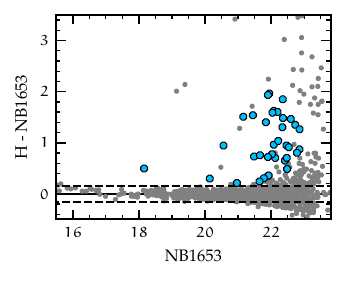}
\end{center} 
\end{minipage}
\caption{The colour--magnitude diagram for HS1700 protocluster at $z=2.30$. The blue points show the emitters with more than 3-sigma colour excesses in K$_{\rm s}$$-$Br$\gamma$, or K$_{\rm s}$$-$NB2167, or K$_{\rm s}$$-$NB1653, respectively. Br$\gamma$ and NB2167 filters can select HAEs at $z=2.30$ while NB1653 can select O3Es at $z=2.30$.
} 
\label{fig:Emitterselections}
\end{figure*}

We select HAEs and \oiii\ emitters at $z=2.3$ by combining NB and broad-band (BB) imaging and making colour--magnitude diagrams (Figure~\ref{fig:Emitterselections}).
The following criteria are used for the selection:
\begin{equation}
  m_{\mathrm{BB,apr}}-m_{\mathrm{NB,apr}}>-2.5\log_{10}\left[1-\frac{\,3\sqrt{\sigma_{\mathrm{BB,apr}}^2+\sigma^2_{\mathrm{NB,apr}}}}{f_{\mathrm{NB, apr}}}\right],
\label{eq:SNcut}
\end{equation}
where $m_{\mathrm{BB,apr}}$ and $m_{\mathrm{NB,apr}}$ is an NB and BB 1.2 arcsec diameter aperture magnitude respectively, $f_{\mathrm{NB, apr}}$ is NB flux density, and flux density of photometric error $\sigma$ in the equation corresponds to a 1-sigma limiting magnitude.
In this work, we do not consider the colour term because \cite{Daikuhara2024} shows that the continuum variation does not affect our conclusion ($< 0.03$) for all NB emitters at $z\sim2$.

We also apply colour cuts to separate true line emitters from contaminants with the following formula:
\begin{equation}
    m_{\mathrm{BB,apr}}-m_{\mathrm{NB, apr}}>0.15.
\end{equation}
These colour cuts correspond to the equivalent width (EW) cuts of emission lines corresponding to rest-frame EW $\sim$ 20 \text{\AA}.
These criteria are determined by the scatter of negative values of $m_{\mathrm{BB, apr}}-m_{\mathrm{NB, apr}}$ with $17<m_{\mathrm{NB,apr}}<20$.

\subsection{Colour -- colour diagram}\label{sec:Colour2}

\begin{figure*}
\begin{minipage}{\columnwidth} 
\begin{center} 
\includegraphics[width=\columnwidth]{Figure/selection/GJK.pdf}
\end{center} 
\end{minipage} 
\begin{minipage}{\columnwidth} 
\begin{center} 
\includegraphics[width=\columnwidth]{Figure/selection/URK.pdf}
\end{center} 
\end{minipage}
\begin{minipage}{\columnwidth} 
\begin{center} 
\includegraphics[width=\columnwidth]{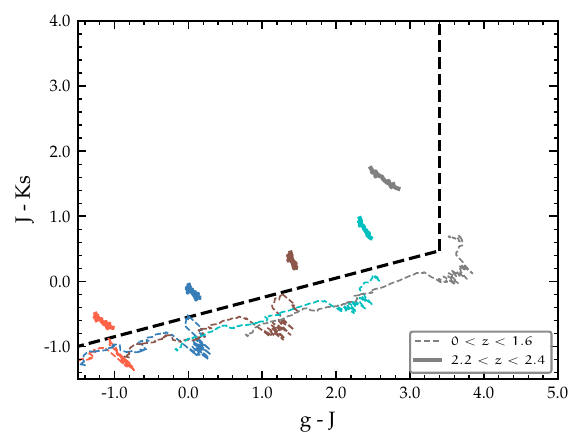}
\end{center} 
\end{minipage} 
\begin{minipage}{\columnwidth} 
\begin{center} 
\includegraphics[width=\columnwidth]{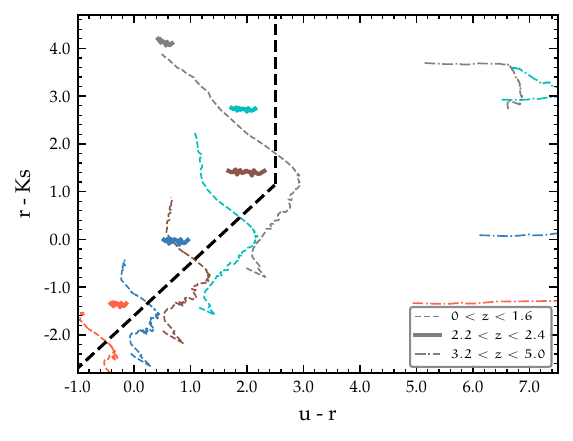}
\end{center} 
\end{minipage}
\caption{The top panels show the model tracks on the colour -- colour diagrams. The solid lines show the galaxies at $2.2<z<2.4$. The dashed lines show the low-$z$ galaxies  ($0<z<1.6$) and the dashed and dotted lines show the high-$z$ galaxies ($3.2<z<5.0$). We set the boundaries so that we can pick out emission line galaxies at $z=2.3$ with reference to these model tracks and spectroscopic data (black dashed line). These lines are derived from model templates of star-forming and quiescent galaxies available in the EAZY code \citep{Brammer2008}. The bottom panels show our NB emitters. Circles represent the selected emitters at $z=2.3$. Grey circles represent the emitters at different redshifts. 
The enclosed squares are confirmed emitters from spectroscopy or paired NB filters.} 
\label{fig:Colorselections}
\end{figure*}

Once we identify line emitter candidates, the next step is to separate those emitters into different lines at different redshifts.
For this purpose, we use either spectroscopic redshifts, if available, or colour -- colour diagrams (Figure~\ref{fig:Colorselections}).

\begin{figure*}
\includegraphics[width=\textwidth]{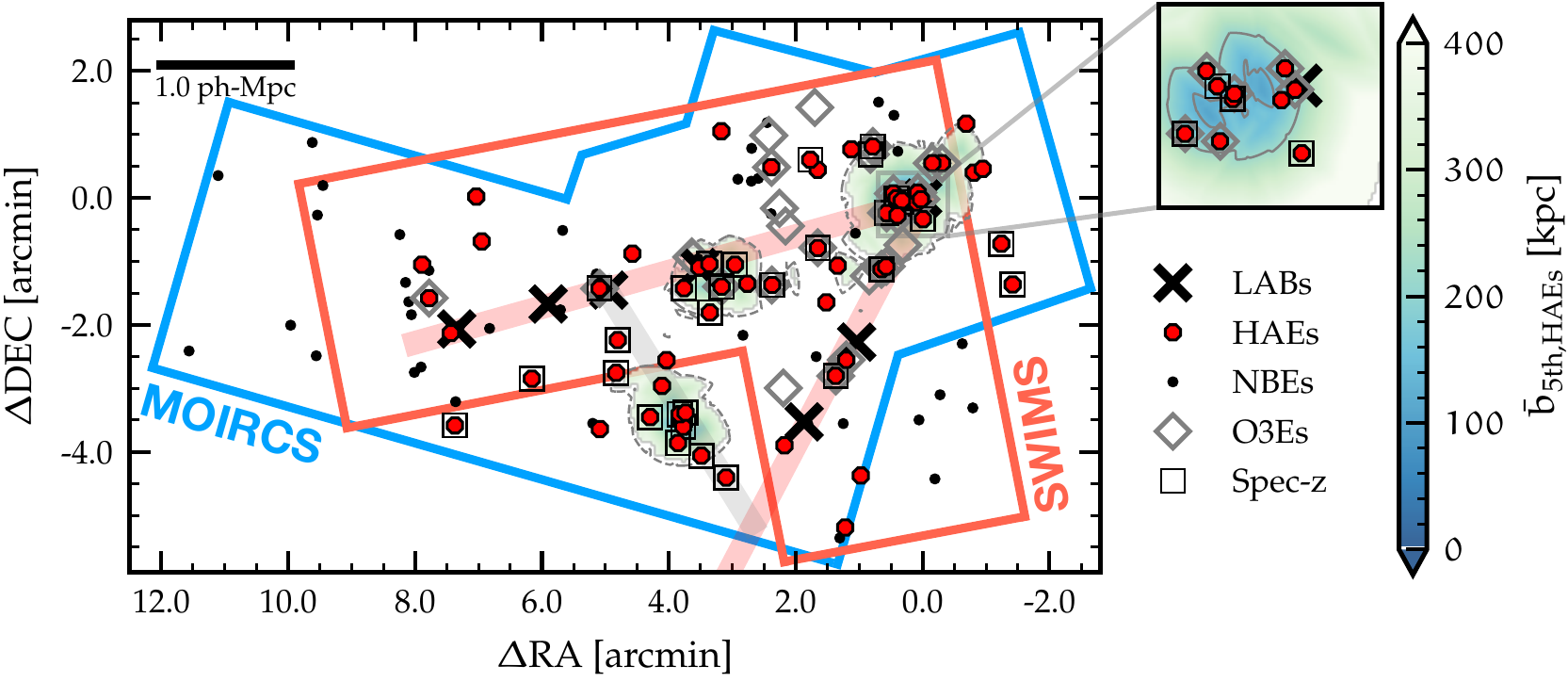}
\caption{The sky coverages of our MOIRCS and SWIMS imaging observations of the protocluster HS1700 overlaid on the spatial distributions of NB-selected emitters. A larger figure of the spatial distribution of galaxy populations will appear in Figure~\ref{fig:dist_hs1700}. Blue and pink lines enclose our MOIRCS and SWIMS coverages, respectively.
The background colour map represents the mean projected distance (equivalent to local density) of HAE galaxies ($\bar{b}_{\mathrm{5th,HAEs}}$), with values ranging from 0 to 400 kpc.
Black cross marks represent LABs.
HAEs and O3Es are marked with red points and grey open diamonds.  Other symbols represent black points for NBEs (HAE, O3E, Pa$\beta$ emitters, etc.), and open squares for spectroscopic confirmed emitters.
The minimum $\bar{b}_{\mathrm{5th,HAEs}}$ corresponds to 76.7 kpc, located at RA = 17:00:48.678, Dec = +64:15:41.96.
}
\label{fig:fov}
\end{figure*}

The HAEs at $z$ = 2.3 in HS1700 are selected so as to satisfy the following two criteria;
\begin{equation}
(g - J) < 3.4,  \quad(J - Ks) > 0.3\times(g - J) - 0.55,\label{eq:colour_1}
\end{equation}
\begin{equation}
(u - r^{\prime}) < 2.5, \quad (r^{\prime} - Ks) > 1.1\times(u - r^{\prime})-1.6.\label{eq:colour_2}
\end{equation}
These criteria are set from spectral models of galaxies and spectroscopic redshift data in such a way that we can maximize the efficiency of separation.
These model colour are derived from model templates of star-forming and quiescent galaxies in EAZY code \citep{Brammer2008}.
The templates (tweak\_fsps\_QSF\_12\_v3) consist of 12 templates derived from the flexible stellar population synthesis code \citep{Conroy2010}.
In the observed colour–colour diagram, there are more galaxies with redder J $-$ Ks colors than predicted by the models.
These objects are likely reddened due to the effects of dust attenuation.
Regarding the spectroscopic data, all but one object satisfy this boundary.
As a result, we identify in total 62 HAEs and 28 \oiii\ emitters associated with the protocluster at $z = 2.3$ (see Figure~\ref{fig:fov}).

If an emitter has a spectroscopic redshift that is consistent with the cluster redshift ($2.28<z<2.32$), we determine that emitter as a member galaxy.
Then, if a galaxy is classified as an emitter with multiple emission lines with multiple NB filters, we also identify it as a member galaxy.
For the other emitters, we apply these two colour-colour diagrams (Figure~\ref{fig:Colorselections}) to classify those emitters and select our target emitters associated to the protocluster.

\subsection{Spectral energy distribution fitting}

For all the confirmed member galaxies and the candidates, we apply SED fitting with all the available photometry in various bands (u, g, r, J, J1, J2, H and Ks) to estimate their stellar masses ($M_{\star}$) and dust attenuations using the {\it CIGALE} code \citep[Code Investigating GALaxy Emission, ver. 2025.0,][]{Burgarella2005,Noll2009,Boquien2019}.
We assume a delayed exponentially declining star formation history ($\mathrm{SFR}(t) \propto t \cdot \exp(-t/\tau)$), \cite{Chabrier2003} IMF, the \cite{Calzetti2000} redding curve, the fixed metallicity of $Z=0.004$, and the nebular to stellar extinction ratio of one.
The e-folding timescale $\tau$ is allowed to vary between $10^9$ and $10^{10}$~yr, and the stellar population age is allowed to vary between $10^{7.6}$ and $10^{9.5}$~yr. The stellar extinction, $A_V$, is allowed to range from 0 to 3~mag.
The redshift is fixed to the spectroscopic one if available, or fixed to $z=2.3$ if not.
We do not include AGN components in our SED models.
When the AGN component is not included in the fitting, the stellar mass tends to be overestimated because a part of the AGN continuum is misinterpreted as stellar emission. 
The derived dust attenuation can also be biased, as the AGN contribution to the optical–infrared SED may be attributed to dust-reprocessed stellar light, generally leading to an overestimation.
We note that the SED fitting in this study is applied only to the HAEs, and not to the AGN candidates (O3Es excluding dual emitters, see Section~\ref{subsec:AGN}).

\section{Results}\label{sec:resutls}

\subsection{Star formation activities in the structures}
\label{subsec:SF}

\subsubsection{Star formation rates}
\label{subsubsec:SFR}

The advantage of NB imaging is that it provides an emission line flux of each object by imaging alone without spectroscopy \citep{Bunker1995}.
We measure the \ha\ emission line flux using the following equation:
\begin{equation}
    F_{\mathrm{Line}} = \frac{f_{\mathrm{NB}}-f_{\mathrm{Ks}}}{1-\Delta_{\mathrm{NB}}/\Delta_{\mathrm{Ks}}}\Delta_{\mathrm{NB}},\label{eq:lineflux}
\end{equation}
in which $f_{\mathrm{NB}}$ and $f_{\mathrm{Ks}}$ are the NB and Ks flux densities, respectively, and $\Delta_{\mathrm{NB}}$ and $\Delta_{\mathrm{Ks}}$ are the FWHMs of NB and Ks filters, respectively.
We need to keep in mind that a color term exists since the effective wavelengths differ slightly between NB and BB. However, we ignore this color term in this work because it is below $0.1$ mag \cite[see][]{Daikuhara2024}.
Br$\gamma$ and NB2167 filters simultaneously enters \ha\ and \nii\ lines.
To remove the \nii\ emission line contribution, we utilize typical mass -- metalicity relation and N2 index of \citet{Steidel2014} following \citet{Daikuhara2024}.

The SFR ([$\mathrm{M_{\odot}\,yr^{-1}}$]) is derived from the formula of \cite{Kennicutt2009}:
\begin{equation}
\mathrm{SFR_{H\alpha}}=4.82\times10^{-42}\left(\frac{L_{\mathrm{H\alpha}}}{\mathrm{erg\,s^{-1}}}\right).\label{eq_SFRha}
\end{equation}
We account for the conversion factor from the Salpeter IMF \citep{Salpeter1955} to the Chabrier IMF \citep{Chabrier2003}, which we assume to be 1.64 \citep{Madau2014}. Dust corrections are conducted based on SED fitting. 

\begin{figure}
\begin{center}
\includegraphics[width=\columnwidth]{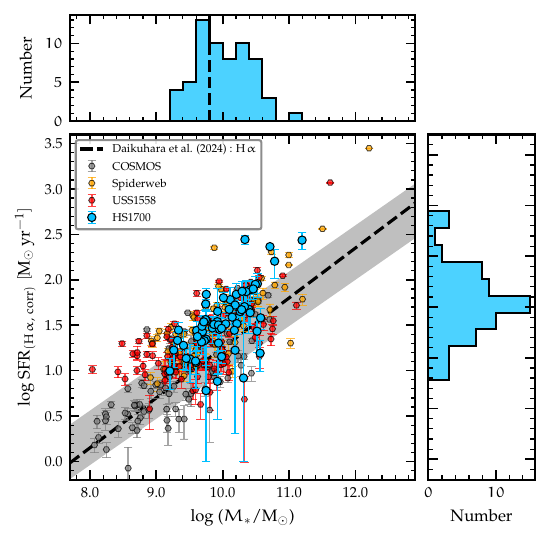}
\end{center}
\caption{The dust-corrected SFR -- $M_{\star}$ relation (star-forming main-sequence diagram) of HAEs in three protoclusters, namely, HS1700 (blue; this work), USS1558 \citep[red;][]{Daikuhara2024}, Spiderweb \citep[orange;][]{Daikuhara2024}, and those in the general field COSMOS \citep[blue;][]{Daikuhara2024}. The upper panel shows a projected histogram of stellar masses, while the right panel shows a projected histogram of dust-corrected SFRs.
}
\label{Fig:hs1700_msandmassfunc}
\end{figure}

Figure~\ref{Fig:hs1700_msandmassfunc} shows the dust-corrected SFR -- $M_{\star}$ relationship for HAEs in the HS1700 protocluster.
Compared to the dust-corrected SFR -- $M_{\star}$ of USS1558 and the Spiderweb \citep{Daikuhara2024}, HS1700 does not show significant differences.
It should be noted that almost all analyses are performed using the same methods (SFR and SED fitting), except for emitter selection due to different limitations of each dataset.
In current observations of HS1700, we do not detect any massive HAE with $M_{\star}/\mathrm{M_{\odot}}>10^{11.5}$.
Note, however, that this does not mean that massive galaxies are absent because previous studies have identified massive DRGs and dusty SFGs exceeding this mass in this protocluster.
The absence of such very massive HAEs in HS1700 suggests either that the system is still in early cluster formation stage to produce such massive galaxies or that we do not see emitters with the amount of dust contents.

\subsection{Spatial distribution of HS1700 protocluster}\label{subsec:filament}
\subsubsection{Filamentary structures of the protocluster}

As described in Section~\ref{sec:Introduction}, the HS1700 protocluster shows unique linear filamentary structures traced by 7 LABs \citep[Umehata et al. in prep;][]{Erb2011}.
We now trace structures of the protocluster using more general star-forming galaxies, namely, HAEs and O3Es and investigate the relationship between star formation activities and the environment (structures) in detail.
Figure~\ref{fig:fov} shows the spatial distribution of HAEs (red points) and O3Es (open orange diamonds). 
Approximately half of the HAEs have been spectroscopically confirmed as member galaxies of the protocluster \citep[28/62$\sim 45\%$;][]{Rudie2012,Steidel2014,Strom2017}. 
We measure the mean projected distance $\bar{b}_{\mathrm{5th}}$ from the fifth nearest HAE to quantify the local environment based on the projected 2-D number density of galaxies.

The densest region of HAEs and O3Es is located at the intersection of the two linear filaments traced by the seven LABs.
The numbering of the LABs is shown in Figure~\ref{fig:hs17000_def_filament}.
The most massive HAE is found around the intersection ($M_{\star} = 10^{11.2} \ \mathrm{M_{\odot}}$). 

We also identify two groups of HAEs, one right on the upper filament and a filamentary structure of HAEs in between the two filaments. 
The former HAE group links with the filament and LAB7.
Another protocluster, SSA22 at $z = 3.1$, also confirmed a clear connection between large-scale filamentary structures and giant LABs \citep{Steidel1998,Steidel2000,Hayashino2004,Yamada2012,Umehata2019}.
The latter HAE group is aligned in a nearly linear configuration, bridging the two LAB filaments.
\citet{Bolda2024} observed two UV continuum-selected galaxies (Q1700-BX710 and Q1700-BX711) in HS1700 using Keck Cosmic Web Imager (KCWI). 
They suggest a new filamentary structure, which is aligned to the elongation of UV continuum-selected galaxies.
The elongations are also confirmed by our HAE and spectroscopy \citep{Rudie2012,Steidel2014,Strom2017}.
\lya\ halos of Q1700-BX710 and Q1700-BX711 are elongated along the galaxy filament identified HAEs or UV-continuum selected galaxies.
This indicates the existence of the third linear filament hosting this group of HAEs, which is drawn by a grey thick transparent line in Figure~\ref{fig:fov}. 
Here we adopt the filament angle as defined by \cite{Bolda2024}, and we consider this third filament as well. 
 
The spatial deviation of the LABs with the same redshift from the best-fit linear function is less than 50 kpc with a median deviation of 20 kpc (Figure~\ref{fig:hs17000_def_filament}), indicating remarkable straight alignment suggestive of a filamentary structure \citep[see also,][]{Erb2011}.
This finding provides compelling evidence that these structures are not merely projection effects but indeed trace the underlying 3D distribution of matter in this region. 
Furthermore, the nearly perpendicular orientation of these filaments with respect to our line of sight offers a unique opportunity to examine their cross-sectional structure.  
This perspective enables us to investigate the internal distribution of galaxies within the filaments and analyze how their properties vary across different regions.  
Such an analysis provides crucial insights into the influence of local filamentary environments on galaxy evolution.  
These regions serve as a valuable laboratory for studying the effects of local filaments on galaxy properties, shedding light on the processes of mass and gas assembly along large-scale structures in the universe.  

\begin{figure}
\begin{center}
\includegraphics[width=\columnwidth]{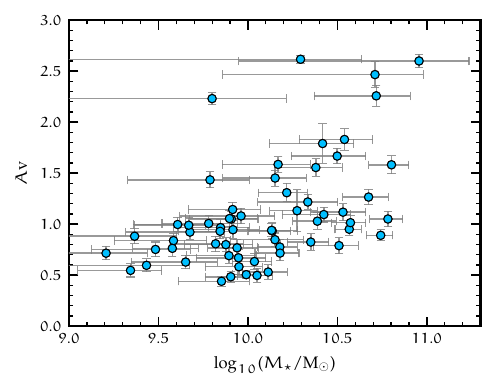}
\end{center}
\caption{The $A_{\mathrm{V}}$ -- $M_{\star}$ relation of HAEs in HS1700. $A_{\mathrm{V}}$ and $M_{\star}$ are derived through SED fitting, and their uncertainties are based on CIGALE. the $A_{\mathrm{V}}$ values span approximately from 0.4 to 2.7.
}
\label{Fig:hs1700_Av}
\end{figure}

\begin{figure*}
\begin{tabular}{cc}
\begin{minipage}{1.1\columnwidth}
\begin{center}
\includegraphics[width=1.02\columnwidth]{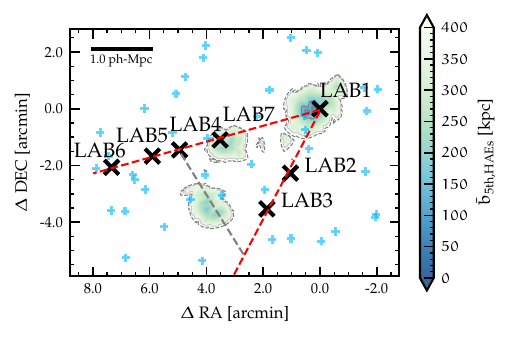}
\end{center}
\end{minipage}
\begin{minipage}{0.9\columnwidth}
\begin{center}
\includegraphics[width=\columnwidth]{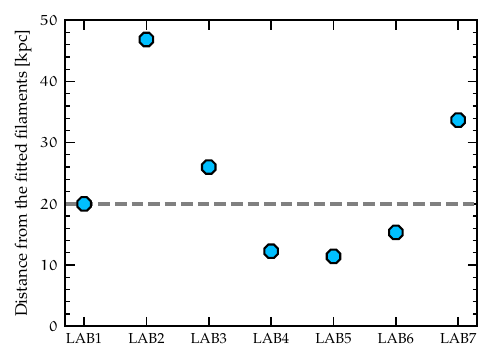}
\end{center}
\end{minipage}
\end{tabular}
\caption{(Left) The numbering of LABs and definitions of the filaments. LAB1 -- LAB6 were identified by \citet{Erb2011}. LAB7 was found by Umehata et al. in prep. We perform a linear fit to LAB1 -- LAB3 (lower filament) and LAB1, LAB4 -- LAB7 (upper filament), respectively, and they are shown by the red dashed lines. (Right) Deviation (distance in kpc) of LABs from the nearest fitted filament. The median value of deviation is only 20 kpc (horizontal dashed line). Given that each LAB extends over $\sim$ 100 kpc, LABs are almost perfectly aligned on these linear filaments.}
\label{fig:hs17000_def_filament}
\end{figure*}

\begin{figure*}
	\includegraphics[width=0.8\textwidth]{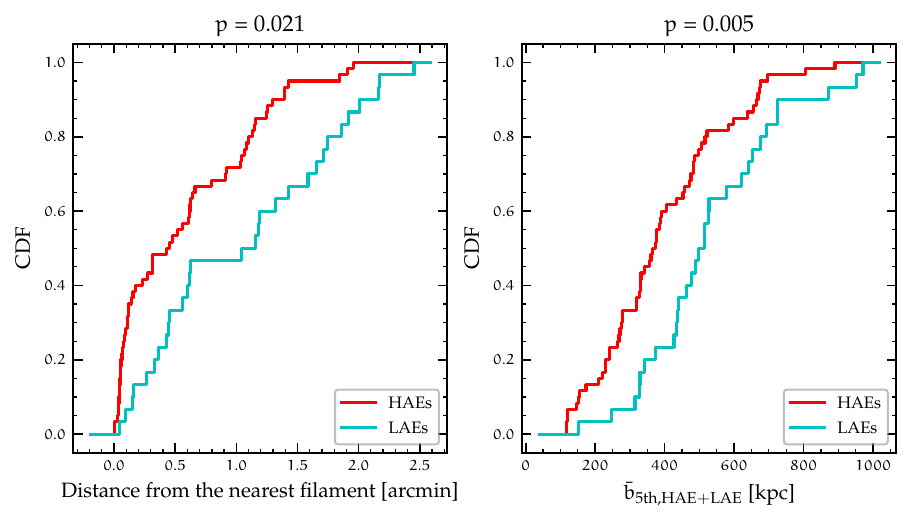}
    \caption{Comparison of spatial distributions of HAEs and LAEs. Red and blue curves show the cumulative distribution functions (CDF) of HAEs and LAEs.
    The left panel shows the distance of each emitter from the nearest filament defined in Figure~\ref{fig:hs17000_def_filament}, while the right panel shows the mean projected distance (equivalent to local density).  A Kolmogorov–Smirnov (K-S) test demonstrates significant differences in the two distributions between HAEs and LAEs on both diagrams ($p=0.021$ and 0.005, respectively). These results suggest that LAEs tend to avoid the high-density regions and filamentary structures where HAEs are highly clustered.}
    \label{fig:ks_dist}
\end{figure*}

\begin{figure*}
\includegraphics[width=\textwidth]{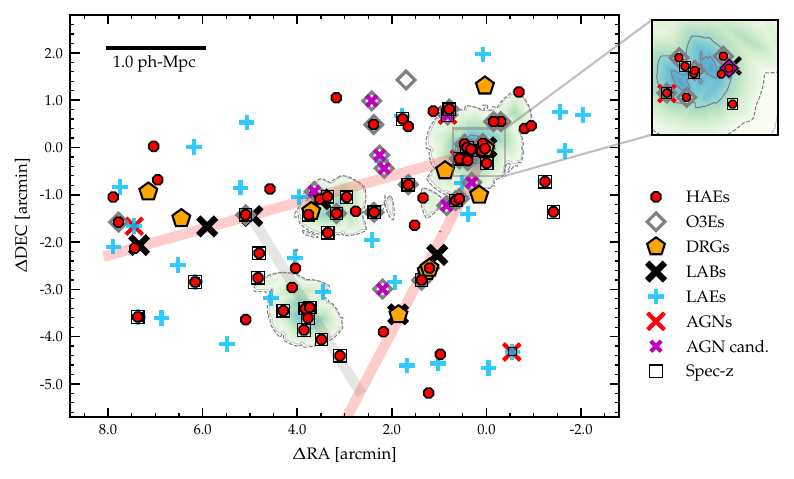}
\caption{Spatial distributions of various galaxy populations, such as HAEs, O3Es,  LAEs, DRGs, AGNs, and AGN candidates in the HS1700 protocluster. 
The background colour map represents the mean projected distance (equivalent to local density) of HAE galaxies ($\bar{b}_{\mathrm{5th,HAEs}}$), with values ranging from 0 to 400 kpc.
LABs are marked with gray crosses. Other symbols represent specific galaxy populations: cyan pentagons for LAEs, orange diamonds for O3Es, black squares for spectroscopically confirmed galaxies, pentagons for DRGs, red octagons for HAEs, magenta crosses for X-ray AGNs \citep{Digby-North2010}. AGN candidates are selected with dual NB images, see in Section~\ref{subsec:AGN}.
The inset zooms into the densest core region at the intersection of the two linear filaments.
This visualization highlights that the densest groups of HAEs are located at the intersection of the two linear filaments. DRGs also tend to be aligned on these filaments. Moreover, all three dense groups of HAEs (including the core) host LABs.
}
\label{fig:dist_hs1700}
\end{figure*}

\subsubsection{Association of distant red galaxies in dense environments}
Distant Red Galaxies (DRGs) are massive galaxies observed at $z>2$ which are characterized by their red colours with $J - Ks > 2.3$ in Vega magnitude or $J - Ks > 1.38$ in AB magnitude \citep{Franx2003,vanDokkum}.
This criterion corresponds to old stellar populations at $z>2$ as predicted by stellar population synthesis models such as  \cite{Bruzual1993}.
The $J-Ks$ colour neatly brackets the 4000\AA\ break feature at $z>2$ which is produced by old stellar populations of ages greater than $\sim$ 1 Gyr.
At the same time, this colour criterion can pick out dusty star-forming galaxies with red UV--optical continua at various redshifts ($z>1$).

\begin{figure*}
  \begin{tabular}{cc}
    \begin{minipage}{\columnwidth}
      \begin{center}
        \includegraphics[width=\columnwidth]{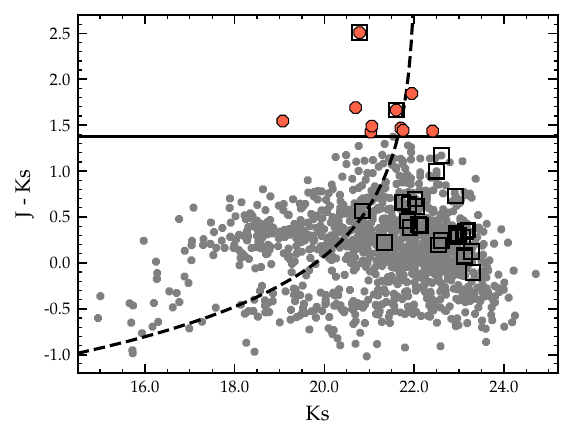}
      \end{center}
    \end{minipage}
    \begin{minipage}{\columnwidth}
      \begin{center}
        \includegraphics[width=\columnwidth]{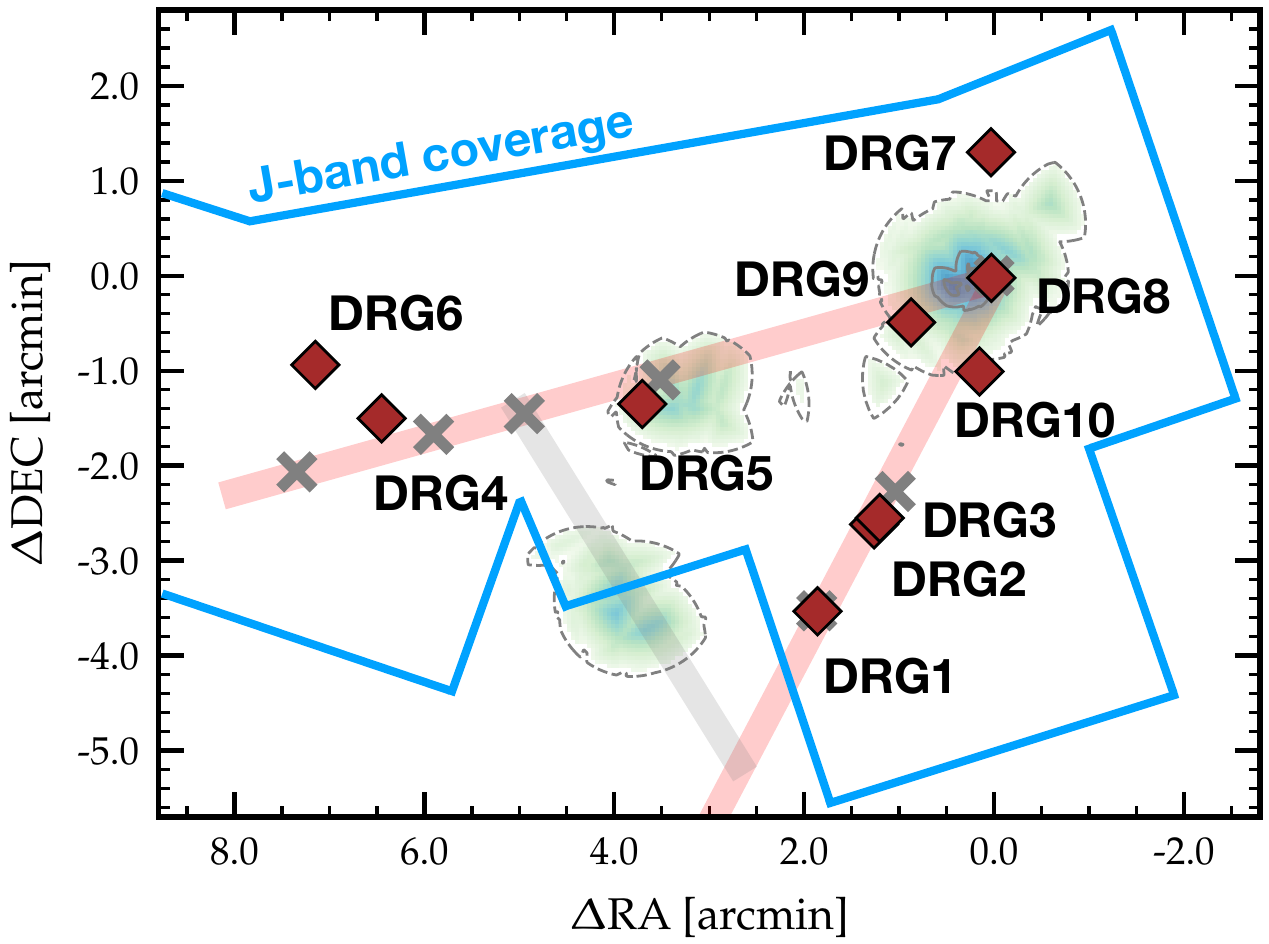}
      \end{center}
    \end{minipage}
  \end{tabular}
\caption{The colour -- magnitude diagram used to select DRGs. The DRGs are defined by J $-$ Ks $>$ 1.38 (solid line). The dashed curve represents a model track for a constant stellar mass of $M_{\star} = 10^{11}\ \mathrm{M_{\odot}}$ assuming passive evolution of stellar populations formed at $z_f$ = 5. We identify 10 massive DRGs in our field. Squares indicate galaxies detected as HAEs, showing that two of the DRGs were also identified as HAEs (DRG3 and DRG8). Both of these DRGs are located near LABs. DRG1 and DRG8 are considered to be associated with the LABs.}
\label{fig:hs1700_DRG}
\end{figure*}

We select 10 DRGs with fixed 1.2 arcsec diameter aperture photometry down to the 3 $\sigma$ limit (Figure~\ref{fig:dist_hs1700}).
Intriguingly, these DRGs tend to be distributed along or near the two linear filaments (Figures~\ref{fig:dist_hs1700} and \ref{fig:hs1700_DRG}).
 DRG1 and DRG8 are considered to be associated with the LABs, based on their spatial coincidence.
DRG3 and DRG8 are also identified as HAEs, and two overlap with LABs (one of them is also a HAE).
Moreover, a DRG (the brightest in Ks) is detected by $^{12}$CO observations and is found to have $z=2.3\pm0.2$ and SFR = 793 $\mathrm{M_{\odot}\ yr^{-1}}$ \citep{Chapman2015}.
Therefore, at least four DRGs are considered to be member galaxies.

To distinguish between quiescent and dusty star-forming galaxies, we examine the Spitzer/MIPS 24 $\mu$m images.
The MIPS data reach a 3$\sigma$ depth of approximately 10 $\mu$Jy \citep{Reddy2010}.
If AGN activity is present, the MIPS 24 $\mu$m flux can also be enhanced by radiation from hot dust surrounding the central AGN.
MIPS-detected DRGs are therefore likely to be either dusty star-forming galaxies or AGNs, although they are still generally massive.
Note, however, that the 24 $\mu$m images are affected by blending issues due to the poor spatial resolution (6 arcsec), which impacts the reliability of the photometry.

\begin{figure*}
	\includegraphics[width=0.8\textwidth]{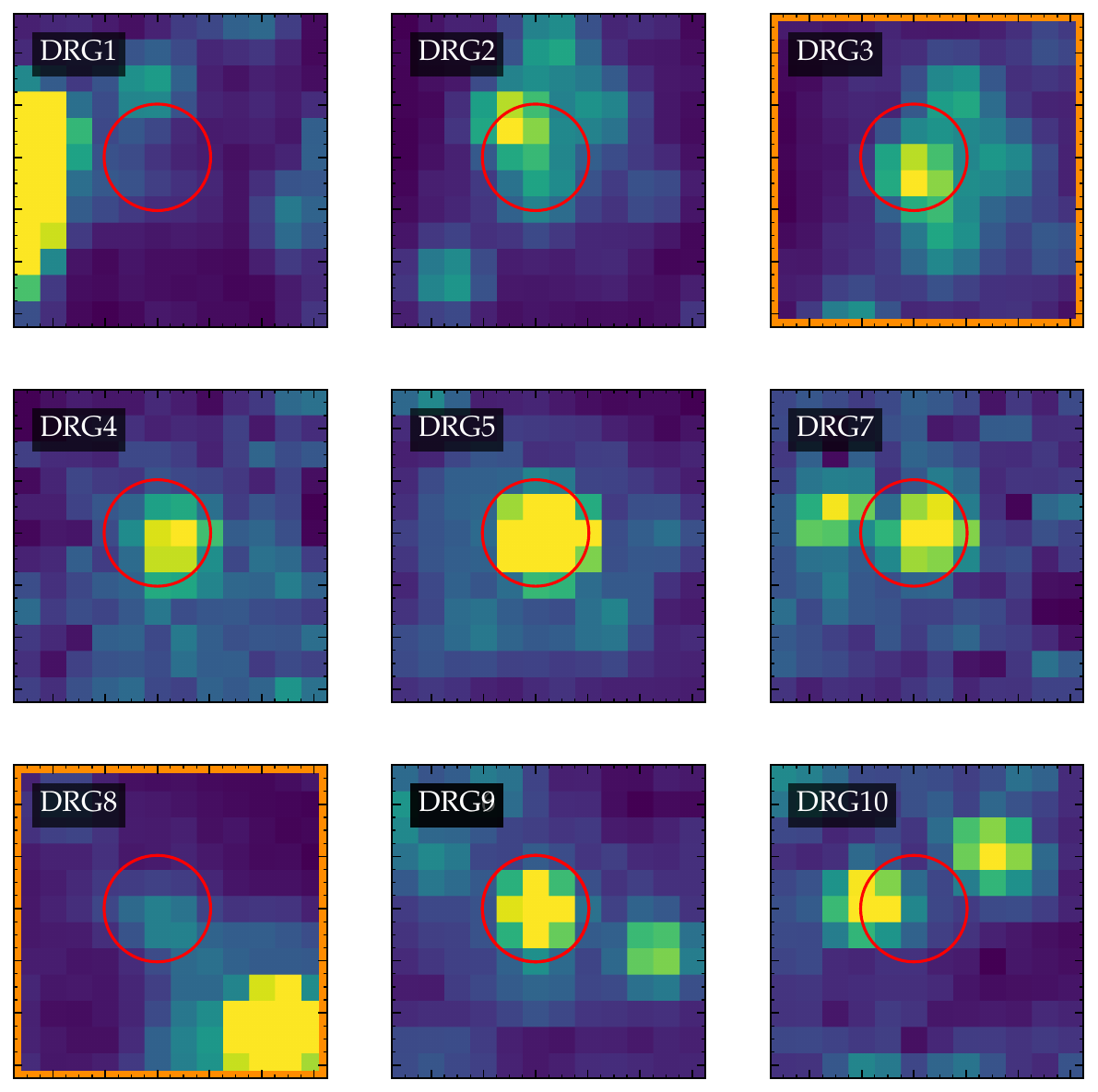}
    \caption{The MIPS images of the DRGs. Box size and the radius of red circles are 30 and 5 arcsec corresponding to 250 and 40 kpc, respectively. The boxes highlighted with orange colour show the HAEs. DRG6 is out renge of MIPS FoV.
    }
    \label{fig:drg_mips_hs1700}
\end{figure*}

We find clear MIPS 24$\mu$m detections for six DRGs (Figure~\ref{fig:drg_mips_hs1700}). 
Four DRGs also have counterparts in the Herschel source catalog, as described by \cite{Kato2016}. 
DRG6 is located outside of the MIPS field of view but is identified as a dusty star-forming galaxy by a Herschel observation \citep{Kato2016}. 
Two DRGs are not detected in the MIP 24$\mu$m band, and they are likely to be quiescent galaxies (QGs).
To confirm their quiescence, we use the medium-band colour J1 $-$ J2 obtained with SWIMS on Subaru, as these adjacent narrower bands can more stringently capture the 4000\AA\ break feature. 
As a result, one QG candidate (DRG10) shows a red colour of J1 $-$ J2 = 0.44, indicating a massive QG with $M_{\star}\sim 10^{11}\ \mathrm{M_{\odot}}$.  
It should be noted that dusty star-forming galaxies, although not detected with MIPS, may exhibit large $J1-J2$ values due to their dust content.
DRG8 exhibits a large $J1-J2$ value of 0.97 but it is identified as an HAE.
See Table~\ref{tab:DRG} for the properties of the DRGs.

\begin{table}
\centering
\caption{Catalog of DRG samples. Ks magnitudes are the Kron magnitude. The colors J $-$ Ks and J1 $-$ J2 are measured using 1.2 arcsec aperture magnitudes. We define DRGs as objects with J $-$ Ks $>$ 1.38. Moreover, the J1 $-$ J2 color traces either the Balmer break or dust reddening; large values of J1 $-$ J2 indicate that an object is either a quiescent galaxy or a dusty star-forming galaxy.}
\label{tab:DRG}
\begin{tabular}{ccccccc}\hline
\ \ \ ID\ \ \ &\ \  Ks \ \ &\ \  J $-$ Ks\ \  &\ \  J1 $-$ J2\ \  &\ \  HAE\ \ \\\hline
DRG1  &  21.95 & 1.84 & 0.08 & no \\
DRG2  &  21.70 & 1.47 & 0.58 & no \\
DRG3  &  21.60 & 1.66 & 0.22 & yes \\
DRG4  &  21.75 & 1.44 & 0.04 & no \\
DRG5  &  19.07 & 1.55 & 0.21 & no \\
DRG6  &  20.70 & 1.69 & 0.36 & no \\
DRG7  &  21.03 & 1.42 & 0.21 & no \\
DRG8  &  20.78 & 2.51 & 0.97 & yes \\
DRG9  &  21.06 & 1.49 & 0.08 & no \\
DRG10 &  22.41 & 1.43 & 0.44 & no \\\hline
\end{tabular}
\end{table}

\subsubsection{Compact star-forming galaxies in the filaments}
\label{subsubsec:CSFGs}

We also investigate the compactness of star formation activity using the same technique described in \citet{Daikuhara2024}.  
Due to differences in data depth, we analyze the properties of this protocluster using a mass-limited sample with \( \log_{10}(M_{\mathrm{star}}/\mathrm{M_{\odot}}) > 10.0 \).  
Note that the SFR range and mass range are not the same as \citet{Daikuhara2024}.

In \citet{Daikuhara2024}, compact star-forming galaxies (CSFGs) are defined as those satisfying \( \mathrm{SFR_{centre}}/\mathrm{SFR_{total}} > 0.4 \), where \( \mathrm{SFR_{centre}} \) and \( \mathrm{SFR_{total}} \) correspond to the SFR within 1.0 arcsec and the total SFR of the galaxy, respectively.  
CSFGs are unique objects inhabiting protoclusters, and the threshold value of 0.4 is adopted since such galaxies are not found in an adjusted field sample of \citet{Daikuhara2024}.  
SFRs are derived from \ha\ line flux measured in NB and BB images (Equation~\ref{eq:lineflux} and Equation~\ref{eq_SFRha}).  
Extended star-forming galaxies (ESFGs) are defined as those with \( \mathrm{SFR_{centre}}/\mathrm{SFR_{total}} < 0.4 \).  
\citep{Daikuhara2024} reported CSFGs are more compact in \ha\ emission than in the continuum. 
This trend is consistent with the compaction scenario, where gas loses angular momentum during galaxy interactions and subsequently falls toward the galaxy center.  

\begin{figure*}
  \begin{tabular}{cc}
    \begin{minipage}{0.50\textwidth}
      \begin{center}
        \includegraphics[width=\columnwidth]{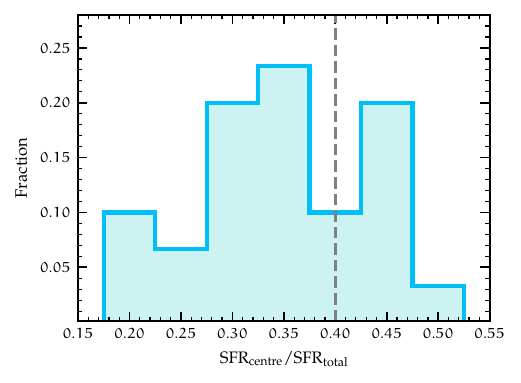}
      \end{center}
    \end{minipage}
    \begin{minipage}{0.50\textwidth}
      \begin{center}
        \includegraphics[width=\columnwidth]{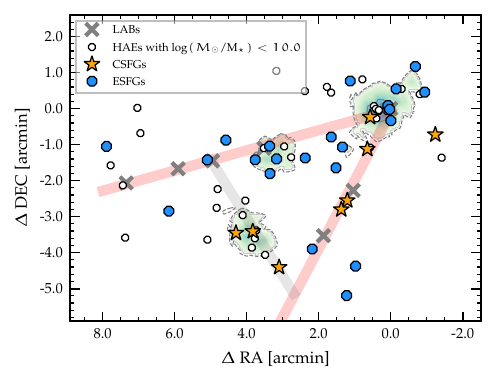}
      \end{center}
    \end{minipage}
  \end{tabular}
\caption{Compact star formation activities in HS1700 protocluster galaxies. The left panel shows the histogram of the limited sample with $\log(M_{\star}/\mathrm{M_{\odot}}>10.0)$. The right panel shows the spatial distribution of compact star-forming galaxies (CSFGs; $\mathrm{SFR_{centre}}/\mathrm{SFR_{total}} > 0.4$), extended star-forming galaxies (ESFGs; $\mathrm{SFR_{centre}}/\mathrm{SFR_{total}} < 0.4$), HAEs with $\log(M_{\star}/\mathrm{M_{\odot}}<10.0)$, and the LABs.
The starmarks, blue filled circles, open black circles and crosses represent CSFGs, ESFGs, HAEs and LABs, respectively.  CSFGs tend to be located in the filaments or in the group of HAEs.}
\label{Fig:hs1700_CSFG_hist}
\end{figure*}

\begin{figure}
\centering
\includegraphics[width=\columnwidth]{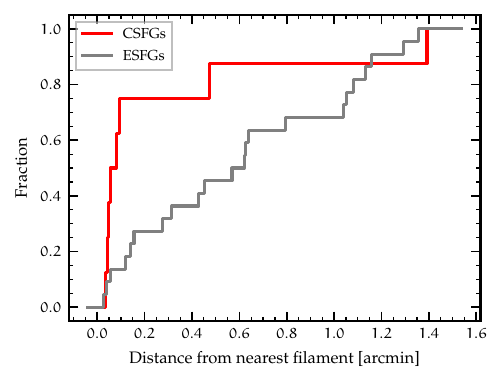}
\caption{The cumulative distribution functions of distance from the nearest filament for CSFGs and ESFGs. A K-S test demonstrates a statistically significant difference in their distributions ($p=0.01$), indicating that the CSFGs tend to be located closer to the filaments.}
\label{fig:CSFGs_KS}
\end{figure}

Figure~\ref{Fig:hs1700_CSFG_hist} shows the distribution of \( \mathrm{SFR_{centre}}/\mathrm{SFR_{total}} \) in HS1700 protocluster, confirming compact star formation activity.  
CSFGs are predominantly located in high-density HAE regions and filaments composed of LAB1 to LAB3 (Figure~\ref{fig:CSFGs_KS}).
The concentration in one of the filaments may reflect the diversity of filamentary environments.

\begin{figure*}
  \begin{tabular}{ccc}
    \begin{minipage}{0.33\textwidth}
      \begin{center}
        \includegraphics[width=\columnwidth]{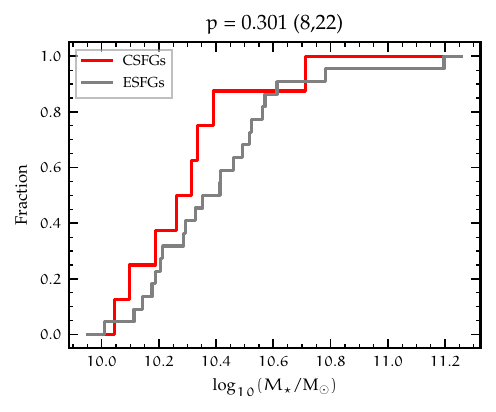}
      \end{center}
    \end{minipage}
    \begin{minipage}{0.33\textwidth}
      \begin{center}
        \includegraphics[width=\columnwidth]{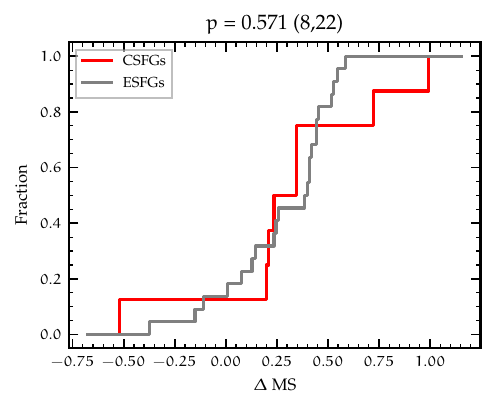}
      \end{center}
    \end{minipage}
    \begin{minipage}{0.33\textwidth}
      \begin{center}
        \includegraphics[width=\columnwidth]{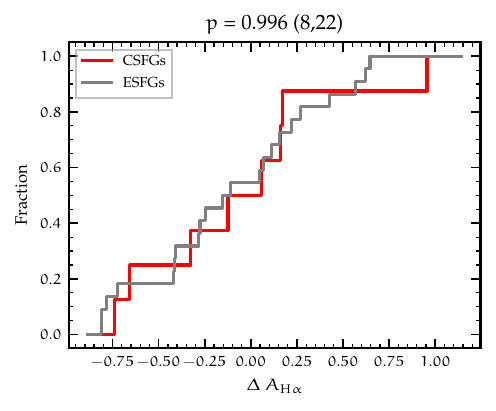}
      \end{center}
    \end{minipage}
  \end{tabular}
\caption{The cumulative distribution functions of stellar mass (left), $\Delta$MS (middle), and $\Delta A_{\mathrm{H}\alpha}$ (right) for CSFG ($\mathrm{SFR_{centre}}/\mathrm{SFR_{total}} > 0.4$), ESFG ($\mathrm{SFR_{centre}}/\mathrm{SFR_{total}} < 0.4$).
Red and gray lines show CSFGs and ESFGs, respectively.}
\label{fig:hs17000_ks_csfgesfg}
\end{figure*}

No significant differences are found in the distribution of star formation activity or dust attenuation between CSFGs and ESFGs (Figure~\ref{fig:hs17000_ks_csfgesfg}).  
In Spiderweb and USS1558, differences in the mass distributions are identified \citep{Daikuhara2024}, but this discrepancy arises from differences in the mass range of the sample.  

The compact nature of star formation in protocluster galaxies are suggested that environmental interactions play a crucial role in redistributing gas and triggering localized starburst activity.  
Moreover, since massive galaxies such as DRGs are thought to originate from galaxy mergers, this result may be linked to their presence along the filaments associated with LABs.  

\subsection{AGN candidates selected with dual narrow-band imaging}
\label{subsec:AGN}

\begin{figure*}
\begin{tabular}{cc}
\begin{minipage}{0.50\textwidth}
\begin{center}
\includegraphics[width=\columnwidth]{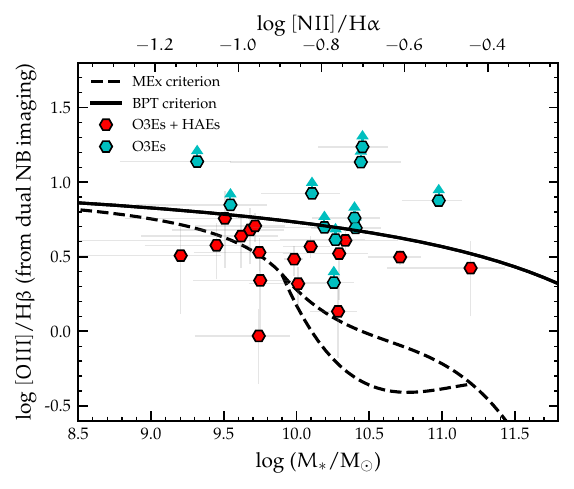}
\end{center}
\end{minipage}
\begin{minipage}{0.50\textwidth}
\begin{center}
\includegraphics[width=\columnwidth]{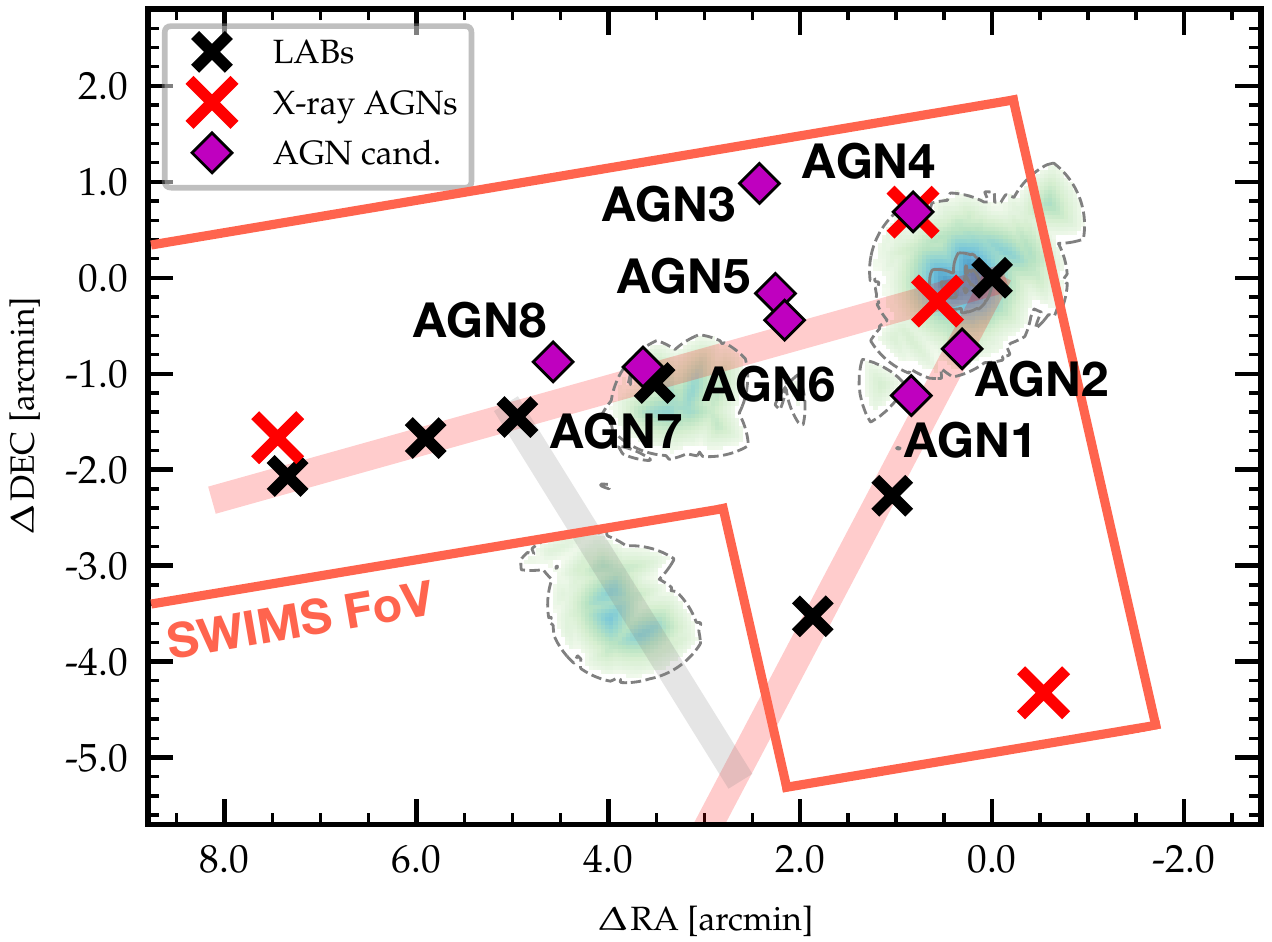}
\end{center}
\end{minipage}
\end{tabular}
\caption{(Left) Mass--Excitation (MEx) diagram showing the relationship between \oiii/\hb\ line ratio and stellar mass for the sample of HAEs and O3Es. The black dashed lines represent the AGNs/SFGs dividing line proposed by \citet{Juneau2011}. The black solid line corresponds to the BPT-based AGNs/SFGs dividing line of \citep{Kewley2001}, assuming a typical mass -- metalicity relation at $z\sim2$ \citep{Steidel2014}. Blue and red points show HAEs and O3Es. We focus on six O3Es with high \oiii/\hb\ ratios (high ionizing states). (Right) Spatial distributions of X-ray AGNs and the AGN candidates selected by the left figure. With the exception of an X-ray AGN, the majority of the objects are again aligned on the filaments or in the outskirt of the protocluster core but avoiding the densest core region.}
\label{Fig:hs1700_MEXBT}
\end{figure*}

\begin{figure*}
	\includegraphics[width=1.0\textwidth]{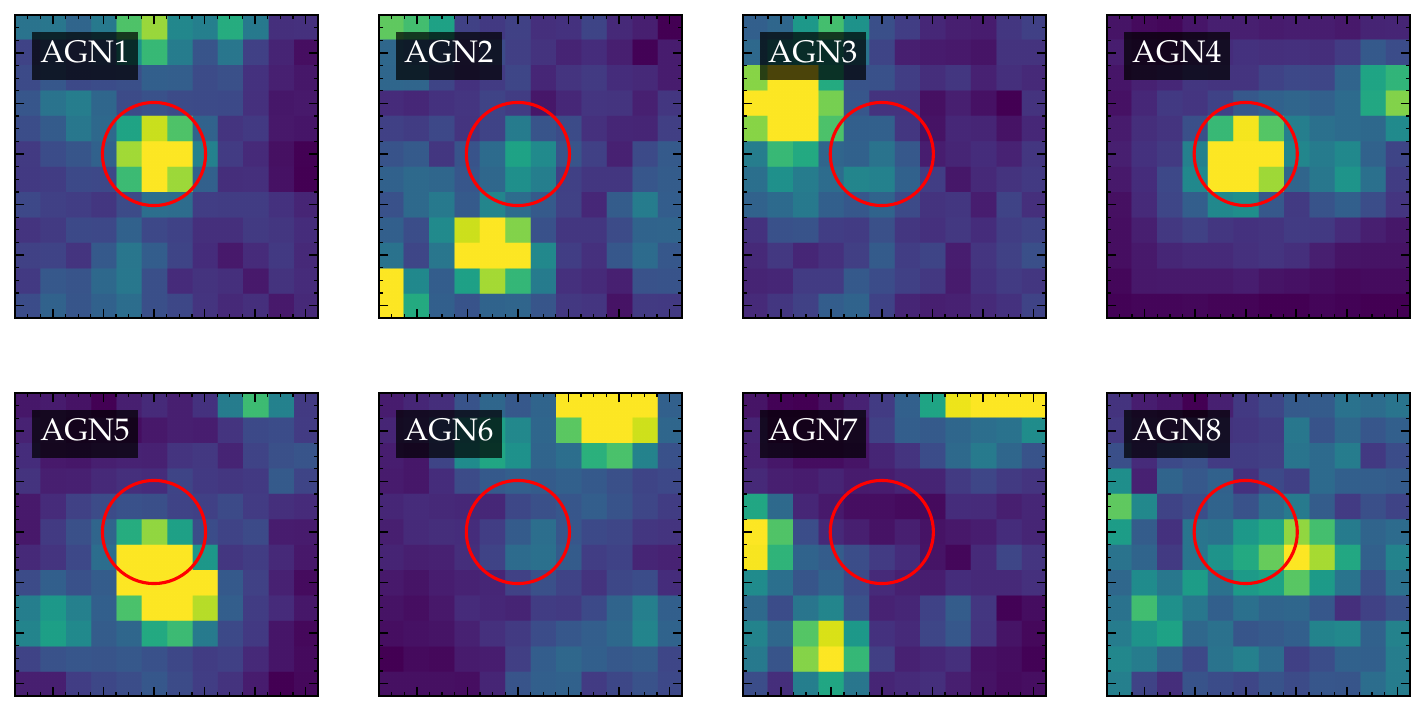}
    \caption{The MIPS images of the AGN candidates. Box size and the radius of the red circles are 30 and 5 arcsec, corresponding to about 250 and 40 kpc, respectively. }
    \label{fig:agn_mips_hs1700}
\end{figure*}

Our coordinated dual NB imaging provides us with the \oiii\ to \ha\ line ratios of individual emitters, which can serve as a good indicator to efficiently select AGN candidates with high ionization states.  
In particular, we focus on O3Es without HAE counterparts, which all lie above the dividing line between AGNs and SFGs shown in Figure~\ref{Fig:hs1700_MEXBT}.  
Using paired NB filters in the NB1653 and BrG(NB2167), we can photometrically derive \oiii/\ha\ ratios once dust attenuation is corrected.
We utilize SED-derived dust extinction value and assume the \citet{Calzetti2000} attenuation curve and case-B recombination (intrinsic \ha/\hb\ line flux ratio is 2.86) at the gas temperature of \( T_e = 10^4 \)~K and an electron density of \( n_e = 10^2 \)~cm$^{-3}$ \citep{Brocklehurst1971}, and convert the \ha\ flux to the \hb\ flux.

Figure~\ref{Fig:hs1700_MEXBT} presents the mass-excitation (MEx) diagram \citep{Juneau2011} used to identify AGN candidates.  
We plot the AGNs/SFGs dividing line based on the BPT diagram \citet{Kewley2001}, assuming a typical mass-metallicity relation (MZR) at $z=2$ from \citet{Steidel2014}.  
The O3E-only objects without \ha\ detections exhibit high enough \oiii/\hb\ ratios that can satisfy both the adjusted BPT-based criterion and the original MEx-AGN criterion \citep{Juneau2011} to be classified as AGNs.  
As a result, eight O3Es are selected as AGN candidates.  
We further examine their MIPS 24$\mu$m images in Figure~\ref{fig:agn_mips_hs1700}. 
AGN1 and AGN4 are clearly detected ($5\sigma$) in MIPS 24$\mu$m.
AGN4 was also selected as an X-ray AGN \citep{Digby-North2010}.
From this, AGN1 and AGN4 may be to be AGNs.
Two X-ray AGN are not detected as AGN candidates with this approach. 
This may be due to small redshift uncertainties that shift their observed properties outside our selection criteria, or to the effects of dust attenuation that obscure emission line features.
To verify AGN candidates, including these two MIPS-detected candidates, we need to examine emission-line diagnostics such as the BPT \citep{Kewley2001} and MEX diagrams \citep{Juneau2011} in the future.

This new method of AGN selection based on paired NB imaging can efficiently select AGN candidates based only on imaging observations without spectroscopy. Although it has not yet been verified by spectroscopic observations, once this technique is established, a large statistical sample of AGNs can be obtained more easily, allowing for more efficient investigation into the relationship between AGNs and their environments along large-scale structures, for example.

AGN candidates (see Figure~\ref{Fig:hs1700_MEXBT} on the right) tend to appear near the protocluster core or along the filaments but avoid the most dense regions of the protocluster core, except for one candidate.
AGN-hosting galaxies are typically massive and may follow different evolutionary pathways compared to SFGs currently residing in the cluster core.  
The AGN activity seems to be associated with intermediate-density regions, where interactions and mergers among massive galaxies are expected most frequently, potentially triggering AGN activities.  
Such interactions can funnel gas toward central supermassive black holes, supplying the fuel necessary to sustain AGN activities.  
To investigate this interesting trend more thoroughly, a larger sample size is required.


\section{Discussion} \label{sec:discussion}
\subsection{Avoidance of LAEs in dense environments and filaments}
\label{subsubsec:LAEs}

We discuss the spatial distributions of more normal and less massive LAEs rather than LABs.
We can see that LAEs tend to avoid overdensity regions and LAB filaments (Figure~\ref{fig:ks_dist}).
To quantify this, we plot the cumulative distribution of LAEs as a function of the distance from the nearest LAB filament and the local density and compare them with the HAE distributions in Figure~\ref{fig:dist_hs1700}.
Here we only consider HAEs with \ha\ NB magnitudes brighter than 23.22, which corresponds to the average 5$\sigma$ limit, because our HAE survey has non-uniform depth.
Also, for the regions where $\Delta$RA $<$ 0 and $\Delta$DEC $>$ 0, we use the distance from the node of the LAB filaments as the nearest distance from the filament, since filamentary structures on this side of the protocluster are not known yet.
As a result, the p-values of the K-S test are only 0.021 and 0.005, respectively, indicating that the two distributions are clearly different statistically.
Note that the possibility of mass segregation affecting the observed trends cannot be fully excluded in this study. 
At least, the dual NB emitters of LAEs and HAEs are located outside the group and filaments of the protocluster.
To further explore this issue, deep NB observations will be required.

\cite{Shimakawa2017b} also shows similar results for a clumpy protocluster USS1558, where LAEs clearly avoid dense protocluster cores traced by HAEs. 
HAEs in overdense regions tend to have much lower escape fractions of \lya\ photons compared to those in lower-density regions. 
This difference in the spatial distributions of HAE and LAE may be due to the larger amount of \hi\ gas or to stronger dust attenuation in higher-density regions such as protocluster cores and surrounding filaments.

\subsection{Cold gas association and accelerated galaxy formation in the protocluster core and surrounding filaments}

First of all, the alignment of LABs along filaments, with the densest HAE regions located at intersections (see Figure~\ref{fig:dist_hs1700}), may provide compelling evidence for cold gas accretion along these structures \citep{Erb2011}. 
The cold gas association along the filaments and the protocluster core is further supported by the lack of normal, low-mass LAEs in these overdense regions compared to outer lower-density regions (Figure~\ref{fig:ks_dist}). 
The \lya\ photons emitted from low-mass galaxies in the filaments and the core are dispersed by resonant scattering by the associated \hi\ gas, and their surface brightness becomes too faint to be detected as LAEs above the sky noise. 
Intrinsically bright massive LAEs can still be visible as diffuse, extended \lya\ emitters, namely LABs.
Moreover, as one of the evidence of the cold gas association, the Q1700-BX711 located in the third filament exhibits a blue-dominated \lya\ emission line, indicative of ongoing cold gas accretion \citep{Bolda2024}. 
Such a blue-dominated \lya\ emission has also been confirmed in another LAB RO-1001 at $z=2.9$, located at the center of a massive galaxy group, supporting the gas accretion scenario based on the morphology and velocity maps of its \lya\ nebula \citep{Daddi2021}. 
However, the absence of bright LABs along this filament (except for its northeastern end) implies that cold gas accretion alone may not be sufficient to produce giant LABs. 
Additional processes, such as intense star formation or AGN activity, may be required to power extended \lya\ emission.

Such gas distribution along the filaments may lead to efficient gas supply to individual galaxies located in the filaments and towards the intersection of the filaments, where the protocluster core develops. 
Such efficient gas accretion may facilitate star formation activities \citep{Keres2005,Dekel2006,Genel2008}. 
In fact, the enhanced star formation at cosmic noon \citep{Hayashi2016,Shimakawa2018uss,Chartab2021,Wang2022,Jose2024,Daikuhara2024}, may be triggered by efficient cold gas accretion. 
In some protocluster at cosmic noon, such efficient cold gas accretion may cause metal deficit of galaxies, including USS1158 \citep{Valentino2015,Li2022,Jose2024}
Furthermore, enhanced star formation has been confirmed only in the protocluster core ($8.9<\log_{10}(M_{\star}/\mathrm{M_{\odot}})<10.2$), while no such enhancement in star formation or metallicity deficit has been observed in the mature protocluster, Spiderweb protocluster $z=2.16$ \citep[$\log_{10}(M_{\star}/\mathrm{M_{\odot}})>9$;][]{Shimakawa2018uss,Jose2023,Daikuhara2024}. 
These results support the scenario of enhanced star formation by efficient cold gas accretion.

Interestingly, the DRGs in HS1700 are also predominantly located along the filaments (Figure~\ref{fig:hs1700_DRG}). 
This finding suggests that in young protoclusters, massive galaxies are forming in an accelerated manner not only in the densest cores but also along the surrounding filaments. 
This trend is consistent with filament study at nearby universe \citep{Chen2017}. 
They reported red galaxy or a high-mass galaxy tends to locate around filaments than a blue or low-mass galaxy, indicating accelerated growth inside the filament.
\cite{Kleiner2017} shows massive galaxies gain \hi\ gas from the intrafilament medium, supporting the cold mode accretion in filamentary environments.
Our result further strengthens the existence of the two filaments traced by LABs, which can also accelerate the formation of massive DRGs in the filaments.

We also find that CSFGs tend to be aligned in the filaments but not particularly in the core.
This may be due to the fact that moderate-density environments are the preferred sites of galaxy mergers/interactions.
Galaxy mergers/interactions are also related to AGN activities and DRGs.
The majority of X-ray AGNs and AGN candidates are also preferentially seen near the filaments and not concentrated in the protocluster core. 
This scenario is also consistent with the alignment of DRGs along the filaments and at the intersection. 
These results suggest that filamentary structures also play important roles in the formation of massive galaxies and AGNs.

The above interpretation also agrees with previous studies that show that moderate overdensities (like filaments) at $z < 2$ promote star formation activity \citep{Smail1999,Koyama2008,koyama2010,Sobral2011,Coppin2012,Darvish2014}.
Moreover, \cite{Darvish2015} reported that star-forming galaxies at $z\sim0.5$ in the filaments are metal-rich compared with the field galaxies.
This work point out that moderate-density regions, including filaments, are important sites for the hierarchical formation of large-scale structures and the formation and growth of galaxies \citep{Coppin2012,Darvish2014,Darvish2015}.

We emphasize the significance of investigating protocluster outskirts from a theoretical point of view.
Semi-analytical models actually highlight the pivotal roles of galaxy groups in galaxy cluster formation \citep{McGee2009}. 
According to these models, approximately 40\% of galaxies in clusters with masses of $10^{14.5} \, h^{-1} \, M_\odot$ at $z = 0$ originate from groups with masses exceeding $10^{13} \, h^{-1} \, M_\odot$. Moreover, galaxies accreted from groups tend to be more massive than field galaxies. 
Examining galaxies within these groups offers crucial insights into the formation of massive galaxies and the environmental influences on their evolution. 

In summary, filaments may serve as the fundamental scaffolding of the high-$z$ universe, channeling cold gas that facilitates galaxy formation and drives early environmental effects.
The surrounding filamentary/clumpy structures of the HS1700 protocluster are the very site of active mass assembly and the best laboratory to investigate the environmentally dependent galaxy formation and evolution.

\subsection{Star formation activity versus environments}
\label{subsubsec:environment}

To investigate the relationship between galaxy properties and protocluster structures, we compare the physical properties of galaxies with distance shorter than $<$ 1 arcmin from the filaments and those outside the filaments. 
We also compare stellar mass, between galaxies in overdense regions ($\bar{b}_{\mathrm{5th}}<400$ kpc) and in other lower-density regions.
We make a comparison of cumulative distributions of $M_{\odot}$, star formation activity, and E(B$-$V) between galaxies located inside and outside the filaments (Figure~\ref{fig:cdf_hs1700_filament}).
The star formation activity is represented by $\Delta\mathrm{MS} = \log_{10}(\mathrm{SFR}) - \log_{10}(\mathrm{SFR_{MS}})$, where $\log_{10}(\mathrm{SFR_{MS}})$ denotes SFR on the star-forming main-sequence (MS);
\begin{equation}
    \log_{10}({\rm SFR}) = 0.51 \log_{10}(M_{\star}/\mathrm{M_{\odot}}) - 3.82,
\end{equation}
based on the results from \cite{Daikuhara2024}.

The width of each filament is tentatively set to 1 arcmin on both sides; however, we confirm that varying this boundary from 0.5 to 1.1 arcmin in 0.1 does not alter the final conclusions as shown in the right panels (Figure~\ref{fig:cdf_hs1700_filament}).
As a result, no significant differences are found in any of these three properties.
A difference in dust attenuation is confirmed if the filament boundary is set to 0.7 or 1.0 arcmin (p-value is less than 0.05), while there is no difference in all the other cases.
However, we summarize that the current data do not provide sufficient evidence for significant differences in dust attenuation between galaxies inside and outside the filaments.
As previously mentioned, dust-attenuated galaxies are less likely to be detected in our observations, suggesting that deeper surveys are necessary for a more thorough investigation of the environmental dependence of dust attenuation.
Although our threshold does not encompass the entire filament, it still captures the high-density regions according to hydro-dynamical simulation \citep[see Figure 13 of][]{GalarragaEspinosa2024}.

We also measure the mean projected distance, $\bar{b}_{\mathrm{5th, HAE+LAE}}$, from the fifth nearest HAEs or LAEs.
We compare galaxies in relatively high-density regions ($\bar{b}_{\mathrm{5th, HAE+LAE}}<400$ kpc) with those in other lower-density regions ($\bar{b}_{\mathrm{5th, HAE+LAE}}>400$ kpc) (see Figure~\ref{fig:dist_hs1700}) but again find no significant differences in the distributions of the stellar mass and star formation activity (Figure~\ref{fig:cdf_hs1700_over}).
A difference in dust attenuation is confirmed if the filament boundary is only set to 420 kpc, while there is no difference in all the other cases.
We conclude that the current data do not clearly support significant differences in dust attenuation between group and outskirts.

We do not detect the environmental dependence of star formation activities.
Some other studies identify an enhancement of star formation activities in low-mass galaxies in the USS1558 protocluster at $z=2.53$ \citep{Jose2024,Daikuhara2024}. or the fraction of SFGs in the filaments is higher than that in the field \cite{Darvish2014}.
The mass range of this study is limited to the massive side, and we do not yet know whether there is an environmental effect on the less massive side of the HS1700.
In this study, the limited depth of the data prevents us from investigating a wide range of stellar masses and SFRs. 
Additionally, dusty galaxies are not selected. 
In future studies, deeper observations will enable us to explore a broader range of stellar masses and SFRs, allowing for a more comprehensive evaluation of enhanced star formation.

\subsection{Origins of LABs}
\label{origin_lab}
The origins of LABs are suggested (1) resonant scattering \citep{Hayes2011}, (2) cold gas accretion \citep{Dijkstra2009,Rosdahl2012,Daddi2022}, (3) AGN-driven fluorescence/photoionization  \citep{Geach2009,Cantalupo2005,Matsuda2011}, (4) Shock \citep{Mori2004,Taniguchi2000}, (5) merger \citep{Yajima2013}, and (6) satellite galaxies \citep{Shimizu2010,Momose2016}.
We confirm filamentary structure by \ha\ and \oiii\ emitters (see Section~\ref{subsec:filament}).
We also identify CSFG associated with filamentary structure (see Figure~\ref{fig:cdf_hs1700_filament}).
These results support that these LABs in the filamentary structure could be explained by (1) resonant scattering, (2) cold gas accretion, (5) merger, and (6) satellite galaxies.
Our dual NB observations or past X-ray observations \citep{Digby-North2010} do not identify any AGN candidates in the LABs.
However, these observations cannot rule out the possibility of contributions from heavily obscured AGNs for (3) \citep{Geach2009,Kim2020}.
The scenario (4) cannot be discussed with the current data set. 
Future deep spectroscopic observations will be key to disentangling these origins.


\section{Summary and conclusions}\label{sec:summary}

We investigate a HS1700+64 protocluster at $z = 2.3$, which exhibits well-defined linear filaments traced by 7 extended LABs, suggesting that it is in an early evolutionary stage and undergoing vigorous assembly. It therefore provides us with an ideal laboratory to investigate the specific influences of filamentary structures on galaxy formation and evolution and the mutual connections between gas accretion and star formation during the early phase of cluster assembly.

Our study utilizes the unique combination of triple NB filters corresponding to \lya, \ha, and \oiii\ emission lines at the protocluster redshift, which allows us to map detailed spatial distributions of various emission line galaxies in and around the protocluster. Moreover, it also provides unique information on emission line ratios of star-forming galaxies from imaging observations only.

We find that HAEs are strongly clustered along filamentary structures and at their intersection. 
This suggests that the filaments serve as gas and matter providers and facilitate the star formation of galaxies therein. 
The galaxies in groups and filaments accumulate along the filaments, merge into the core, and eventually evolve into rich galaxy clusters today. 
The filaments therefore, play crucial roles in galaxy formation and mass assembly of galaxy clusters.

DRGs are predominantly distributed along the filaments, indicating that the formation of massive galaxies is accelerated not only in the dense cores of young protoclusters but also along the surrounding filamentary structures. 
In addition, CSFGs with centrally concentrated star formation tend to reside in high-density regions along the filaments. 
This suggests that galaxy interactions or mergers are also enhanced in these regions.

The dual NB technique enables us to derive emission line ratios, thereby providing valuable information on various properties of the galaxy, such as the association with \hi\ gas (Section~\ref{subsec:filament}) and the presence of AGNs (Section~\ref{subsec:AGN}).

The coordinated dual NB imaging (\ha\ and \lya) indicates that normal less massive LAEs tend to avoid overdense regions and the linear filaments traced by LABs and HAEs (Figure~\ref{fig:ks_dist}). 
We quantify this by comparing the distributions of LAEs and HAEs as a function of local density and distance from the nearest filaments and confirmed this trend quantitatively with a K-S test.
Similar trends were reported in another young, clumpy protocluster USS1558 by \cite{Shimakawa2017b}.
These trends can be interpreted by increased \hi\ gas association and dust attenuation in the protocluster cores and in the filaments, and the \lya\ photons are more scattered or absorbed with respect to \ha\ photons. 
The dust attenuation derived from the SED fitting does not differ significantly between the inner and outer regions of the filaments, nor between the inner and outer regions of groups. 
This suggests that an amount of \hi\ gas may exist in the filamentary or group regions of the HS1700 protocluster.
However, we cannot entirely rule out the impact of mass segregation, which may influence the observed trends. 
To further investigate this issue, deeper NB observations will be necessary.

This work also presents a novel method for selecting AGN using dual NB imaging (\ha\ and \oiii). 
Environmental conditions in protoclusters may also influence AGN activity, potentially through mergers and/or enhanced gas accretion onto supermassive black holes, driven by the concentration of gas toward the galaxy centre. 
We select AGN candidates as objects with high \oiii/\ha\ emission line ratios derived from the dual NB imaging. 
We find that the AGN candidates are preferentially located along the filaments or surrounding the group, but avoiding the densest region.
The preferred emergence of AGN candidates in such intermediate-density environments may indicate a role in triggering AGN activities in the outskirts of the protocluster and in the surrounding filaments. Frequent galaxy interactions and mergers in these regions likely facilitate the accretion of gas onto supermassive black holes.
However, spectroscopic follow-up is required to confirm these candidates and validate the reliability of this selection method. In addition, to draw more definitive conclusions, a larger sample size is necessary.

Future large spectroscopic surveys, such as those conducted with PFS, will make a significant contribution to unambiguously establishing the relationship between AGN activity and environment on a large statistical sample. Furthermore, future observations, particularly with space-based observatories such as JWST, Roman and Euclid, as well as next-generation ground-based facilities like ULTIMATE-Subaru, ELT, and TMT will be crucial for probing the low-mass galaxies in protoclusters and for uncovering the full extent of environmental effects on galaxy formation and evolution. By combining NB imaging, spectroscopic follow-up observations, and theoretical simulations, we can further explore the interplay among gas accretion/ejection, star formation, and structural evolution in these environments, ultimately advancing our understanding of galaxy formation and the role of large-scale structures in the universe.

\section*{Acknowledgements}
This research is based on data collected at the Subaru Telescope, which is operated by the National Astronomical Observatory of Japan. 
We are honored and grateful for the opportunity to observe the Universe from Maunakea, which is cultural, historical, and natural significance in Hawaii.
SWIMS observations at Subaru Telescope was supported by MEXT/JSPS KAKENHI grant No. 20H00171.
This work was supported by JST SPRING, Japan Grant Number JPMJSP2114, JSPS Core-to-Core Program (grant number: JPJSCCA20210003), Graduate Program on Physics for the Universe (GP-PU) Tohoku University and JSPS KAKENHI Grant Numbers 25K23411.
TK acknowledges financial support from JSPS KAKENHI Grant Numbers 24H00002 (Specially Promoted Research by T. Kodama et al.) and 22K21349 (International Leading Research by S. Miyazaki et al.).
HK acknowledges financial support from JSPS KAKENHI Grant Numbers 23KJ2148 and 25K17444.
SK acknowledges financial support from JSPS KAKENHI Grant Numbers 24KJ0058 and 24K17101.
HU acknowledges support from JSPS KAKENHI Grant Numbers 20H01953, 22KK0231, 23K20240, and 25K01039.
JMPM acknowledges that this project has received funding from the European Union’s Horizon research and innovation programme under the Marie Skłodowska-Curie grant agreement No 101106626.
JMPM acknowledges support from the Agencia Estatal de Investigación del Ministerio de Ciencia, Innovación y Universidades (MCIU/AEI) under grant (Construcción de cúmulos de galaxias en formación a través de la formación estelar oscurecida por el polvo) and the European Regional Development Fund (ERDF) with reference (PID2022-143243NB-I00/10.13039/501100011033).
KF acknowledges that this project was supported received funding from the Global-LAMP Program of the National Research Foundation of Korea (NRF) grant funded by the Ministry of Education (No. RS-2023-00301976).

\section*{Data availability}
The Subaru observational data is obtained through the Subaru Mitaka Okayama Kiso Archive (SMOKA) system.
The Spitzer MIPS data used in this work are publicly available via the NASA/IPAC Infrared Science Archive (IRSA). The WHT data are available from the ING Archive.
Additionally, our NB-selected galaxy sample and associated physical quantities are available upon reasonable request by contacting the first author.



\bibliographystyle{mnras}
\bibliography{main} 



\appendix
\section{Extra materials}
\begin{figure*}
\begin{minipage}{\columnwidth} 
\begin{center} 
\includegraphics[width=\columnwidth]{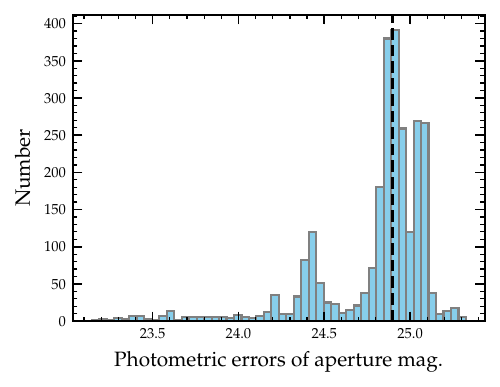}
\end{center} 
\end{minipage} 
\begin{minipage}{\columnwidth} 
\begin{center} 
\includegraphics[width=\columnwidth]{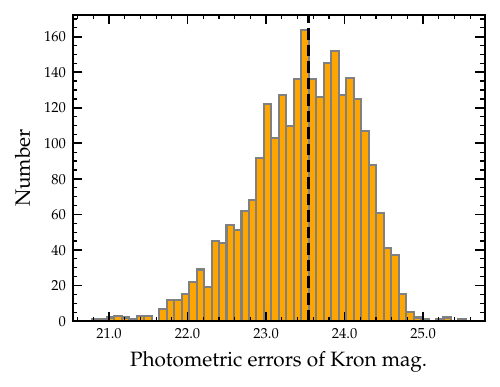}
\end{center} 
\end{minipage}
\caption{The histogram represents the 1$\sigma$ values for each detected object in Br$\gamma$-band mosaic image. The left panel shows the $1\sigma$ error of a 1.2 arcsec diameter aperture magnitude. The right panel shows the $1\sigma$ error of a Kron aperture magnitude. We measure the background level as a function of weight value and aperture size. We plot all detected sources in Br$\gamma$ band image using {\it SExtractor}. The dotted line indicates the median value of photometric error. 
Three peaks can be seen due to the mosaic of the three data sets with different depths.} 
\label{fig:err_1}
\end{figure*}

\begin{figure*}
\begin{minipage}{\columnwidth} 
\begin{center} 
\includegraphics[width=\columnwidth]{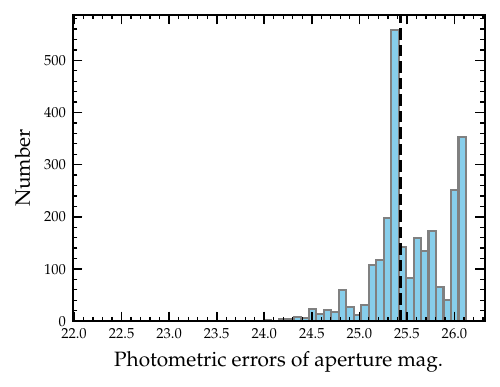}
\end{center} 
\end{minipage} 
\begin{minipage}{\columnwidth} 
\begin{center} 
\includegraphics[width=\columnwidth]{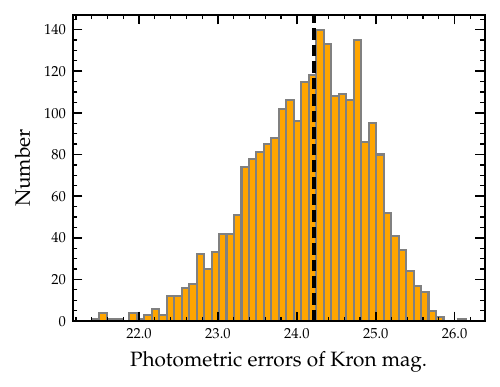}
\end{center} 
\end{minipage}
\caption{The histogram represents the 1$\sigma$ values for each detected object in Ks-band (MOIRCS).} 
\begin{minipage}{\columnwidth} 
\begin{center} 
\includegraphics[width=\columnwidth]{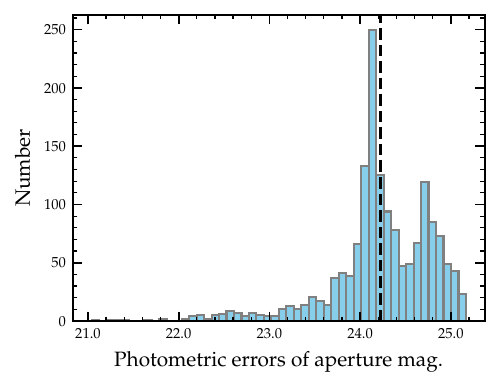}
\end{center} 
\end{minipage} 
\begin{minipage}{\columnwidth} 
\begin{center} 
\includegraphics[width=\columnwidth]{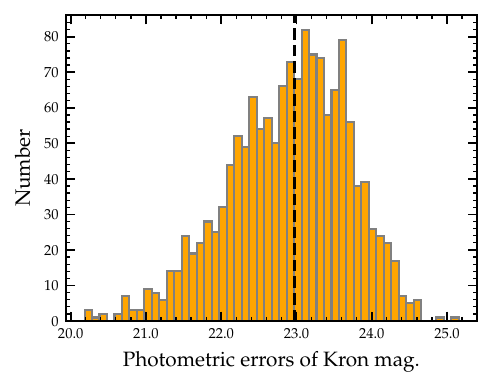}
\end{center} 
\end{minipage}
\caption{The histogram represents the 1$\sigma$ values for each detected object in NB2356-band.} 
\label{fig:err_2}
\end{figure*}

\begin{figure*}
\begin{minipage}{\columnwidth} 
\begin{center} 
\includegraphics[width=\columnwidth]{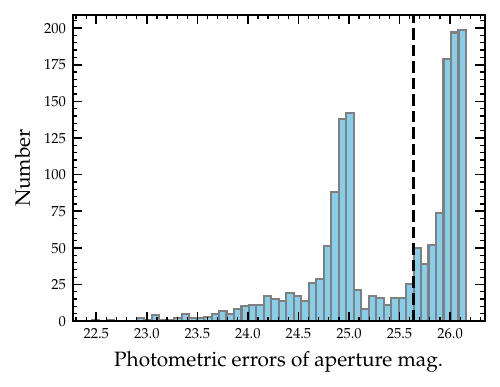}
\end{center} 
\end{minipage} 
\begin{minipage}{\columnwidth} 
\begin{center} 
\includegraphics[width=\columnwidth]{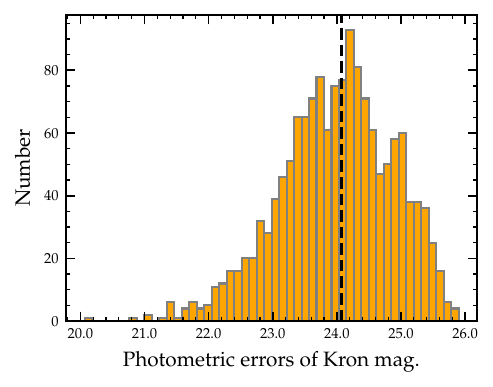}
\end{center} 
\end{minipage}
\caption{The histogram represents the 1$\sigma$ values for each detected object in Ks-band (SWIMS).} 
\label{fig:err_4}
\end{figure*}

\begin{figure*}
\begin{minipage}{\columnwidth} 
\begin{center} 
\includegraphics[width=\columnwidth]{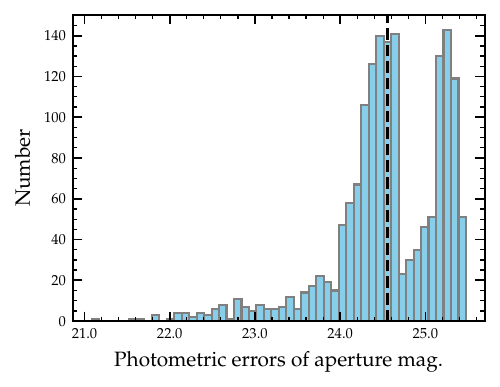}
\end{center} 
\end{minipage} 
\begin{minipage}{\columnwidth} 
\begin{center} 
\includegraphics[width=\columnwidth]{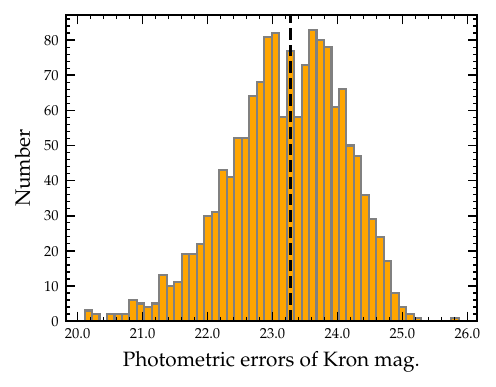}
\end{center} 
\end{minipage}
\caption{The histogram represents the 1$\sigma$ values for each detected object in NB1534-band.} 
\label{fig:err_5}
\end{figure*}

\begin{figure*}
\begin{minipage}{\columnwidth} 
\begin{center} 
\includegraphics[width=\columnwidth]{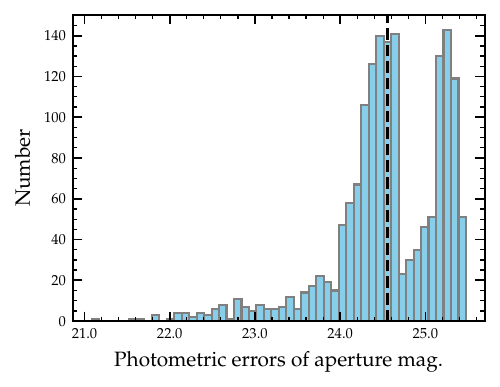}
\end{center} 
\end{minipage} 
\begin{minipage}{\columnwidth} 
\begin{center} 
\includegraphics[width=\columnwidth]{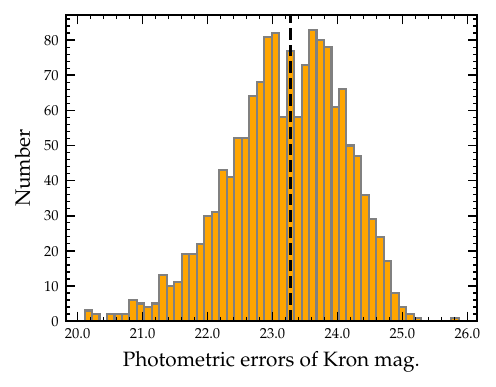}
\end{center} 
\end{minipage}
\caption{The histogram represents the 1$\sigma$ values for each detected object in H-band (SWIMS).} 
\label{fig:err_6}
\end{figure*}

Figures~\ref{fig:err_1} -- Figure~\ref{fig:err_6} show the 1 $\sigma$ values for detected sources in each band. The left panel shows the $1\sigma$ error for a 1.6 arcsec diameter aperture magnitude, while the right panel presents the $1\sigma$ error for a Kron aperture magnitude. The background levels were measured as a function of weight value and aperture size for the detected sources in the Br$\gamma$-band image by {\it SExtractor}.

\begin{figure*}
    \centering
    \includegraphics[width=0.95\textwidth]{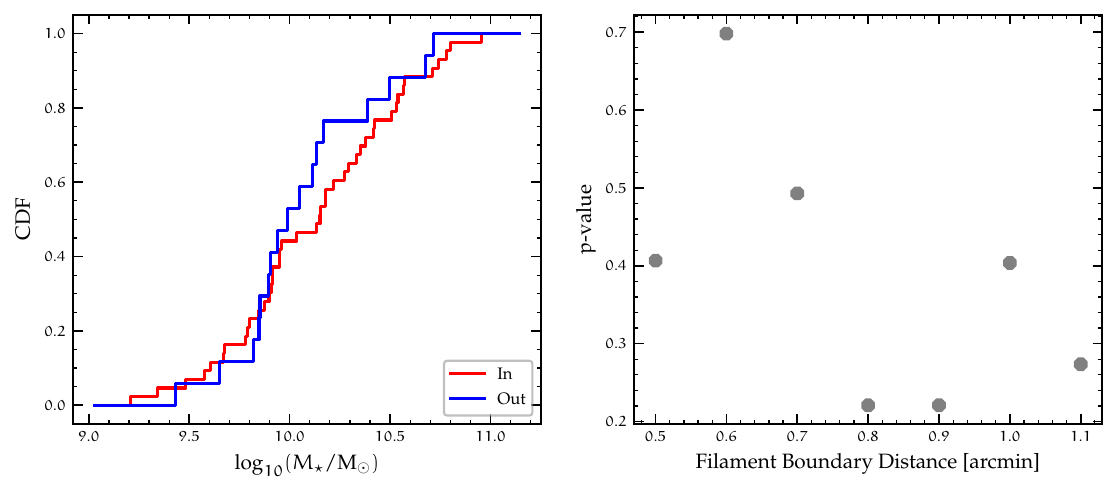}
    \includegraphics[width=0.95\textwidth]{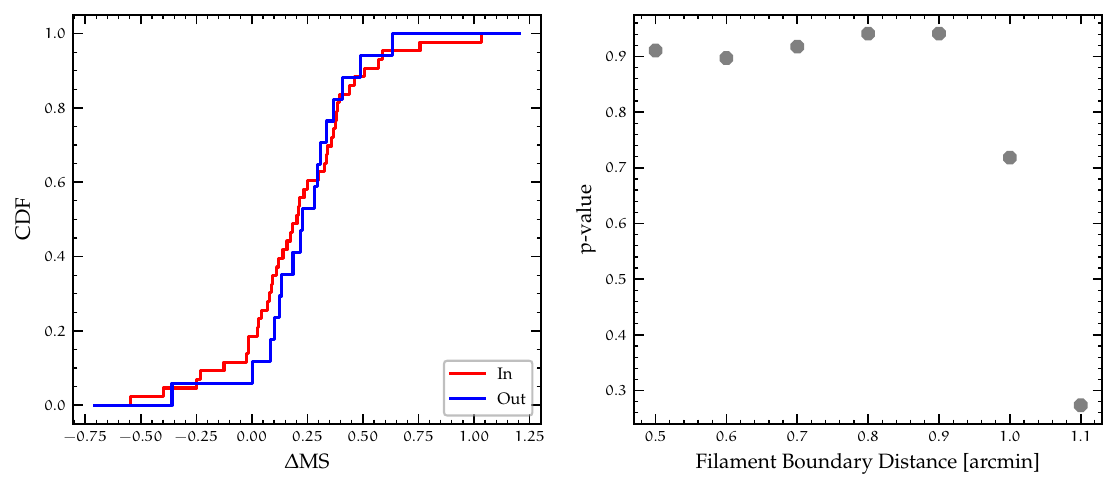}
    \includegraphics[width=0.95\textwidth]{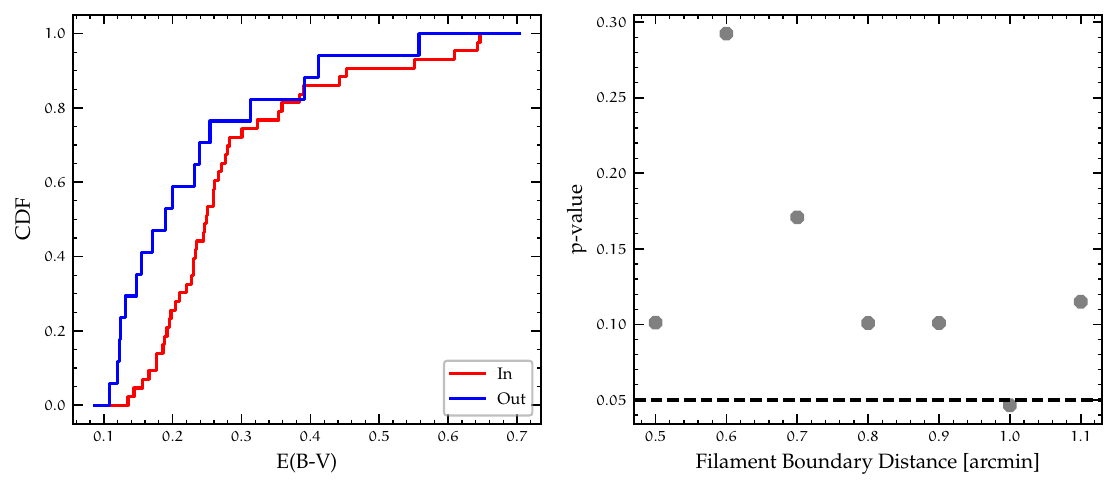}
    \caption{The left panels show the cumulative distribution functions of stellar mass (top), $\Delta$MS (middle), and $\Delta A_{\mathrm{H}\alpha}$ (bottom) for regions inside and outside the filaments. The right panels display how the $p$-values change as the boundary separating the inside and outside of the filaments is varied. The results indicate no significant differences in any of the distributions. While one $p$-value for the dust distribution fall below 0.05, the lack of uniformity suggests that there are no significant differences between the inside and outside of the filaments.}
    \label{fig:cdf_hs1700_filament}
\end{figure*}

\begin{figure*}
	\includegraphics[width=0.95\textwidth]{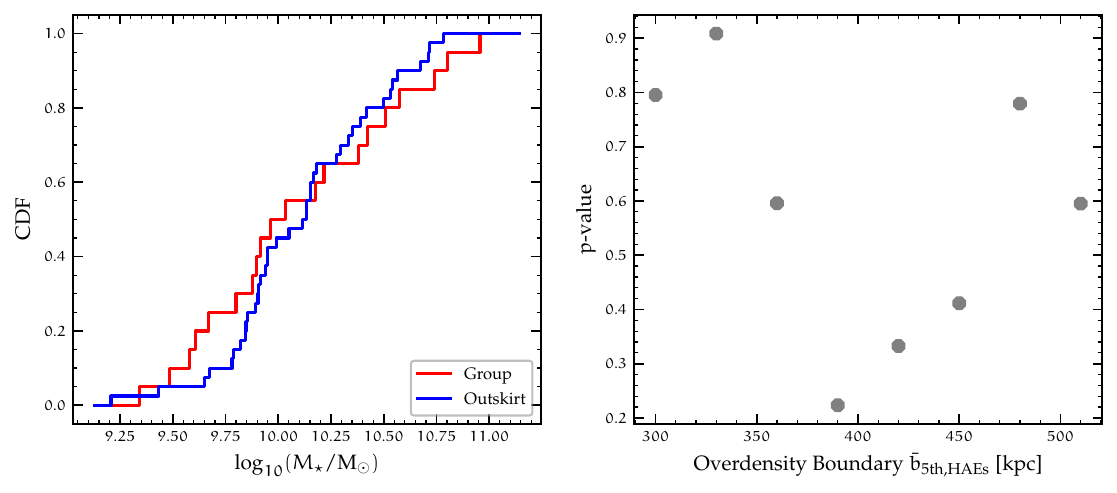}
    \includegraphics[width=0.95\textwidth]{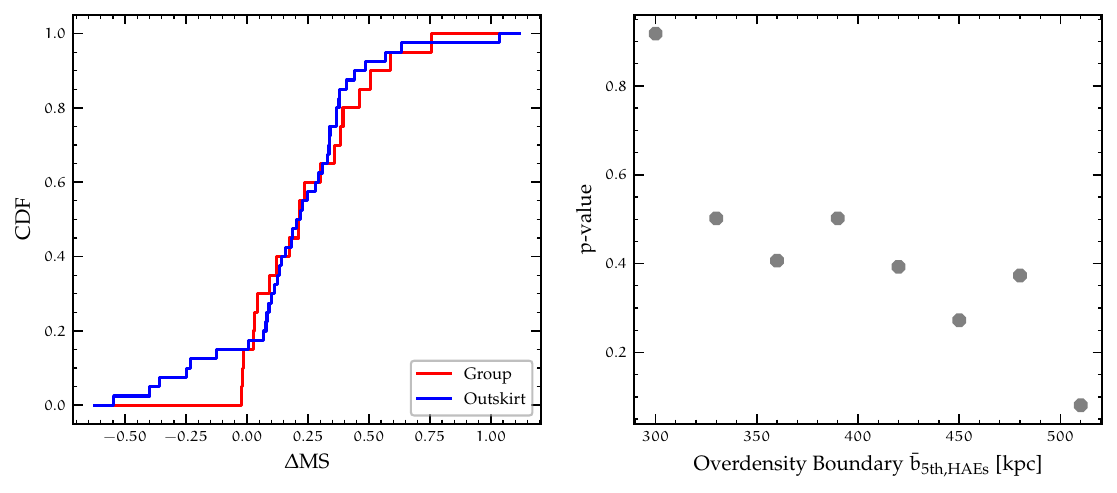}
    \includegraphics[width=0.95\textwidth]{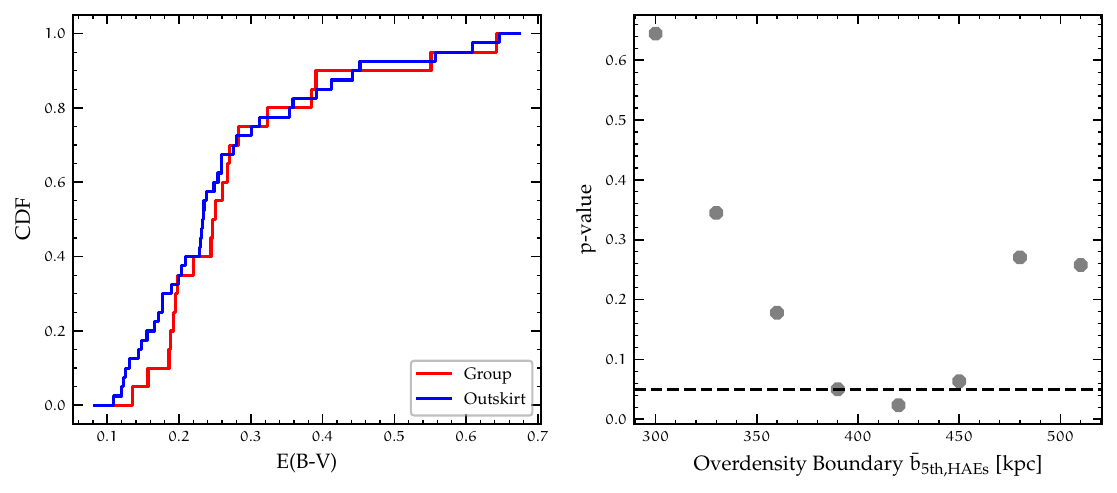}
    \caption{The left panels show the cumulative distribution functions of stellar mass (top), $\Delta$MS (middle), and $\Delta A_{\mathrm{H}\alpha}$ (bottom) for regions inside and outside the group (dense region) of HAEs and LAEs. The red line shows the cumulative distribution function (CDF) in the group ($\bar{b}_{\mathrm{5th,HAEs+LAEs}}<300$ kpc). The blue line shows the CDF in the outskirts of the dense regions ($\bar{b}_{\mathrm{5th,HAEs+LAEs}}>300$ kpc). While one $p$-value for the dust distribution fall below 0.05, the lack of uniformity suggests that there are no significant differences between the proto-cluster group and outskirt.}
    \label{fig:cdf_hs1700_over}
\end{figure*}

Figures~\ref{fig:cdf_hs1700_filament} and Figure~\ref{fig:cdf_hs1700_over} presents the cumulative distribution functions of stellar mass, $\Delta$MS, and $\Delta A_{\mathrm{H}\alpha}$ for regions inside and outside the filaments, along with the variation of $p$-values as the division boundary between these regions is varied. The results indicate no significant differences in any of the distributions.

\bsp	
\label{lastpage}
\end{document}